\documentclass[letterpaper]{article}
\usepackage{fullpage}
\usepackage{graphicx}
\usepackage[sumlimits]{amsmath}
\usepackage{multirow}
\usepackage{subfigure}
\usepackage{url}
\usepackage{varioref}
\usepackage{afterpage}
\usepackage{boxedminipage}
\begin{document}
\title{Repository Replication Using NNTP and SMTP}

\author{Joan A. Smith, Martin Klein, Michael L. Nelson\\
Old Dominion University, Department of Computer Science\\
Norfolk, VA 23529 USA}
\maketitle
\begin{center}
	\{jsmit, mklein, mln\}@cs.odu.edu
\end{center}

\begin{abstract}
We present the results of a feasibility study using \emph{shared,
existing} network-accessible infrastructure for repository replication.
Our goal is not to ``hijack'' other sites' storage, but to take advantage
of protocols which have persisted through many generations and which are
likely to be supported well into the future.
We utilize the SMTP and NNTP protocols to replicate both the metadata 
and the content of a digital library, using 
OAI-PMH and the related Apache web
server module, mod\_oai, to facilitate management of the replication process.
We investigate how dissemination of repository contents can be
piggybacked on top of existing email and Usenet traffic.  Long-term
persistence of the replicated repository may be achieved thanks to current
policies and procedures which ensure that email messages and news posts
are retrievable for evidentiary and other legal purposes for many years
after the creation date.  While the preservation issues of migration and
emulation are not addressed with this approach, it does provide
a simple method of refreshing content with various partners for smaller
digital repositories that do not have the administrative resources for
more sophisticated solutions.
\end{abstract}

\section{Introduction}
We propose and evaluate two repository replication models that rely
on \emph{shared, existing} network-accessible infrastructure. 
Our goal is not to ``hijack'' other sites' storage, but to take advantage
of protocols which have persisted through many generations and which are
likely to be supported well into the future.
The premise is that if
archiving can be accomplished within a widely-used, already deployed
infrastructure whose operational burden is shared among many partners,
the resulting system will have only an incremental cost and be tolerant
of dynamic participation.  With this in mind, we examine the feasibility
of repository replication using Usenet news (NNTP, \cite{rfc:977})
and email (SMTP, \cite{rfc:821}).

There are reasons to believe that both email and Usenet could function as
persistent, if diffuse, archives.  NNTP provides well-understood methods
for content distribution and duplicate deletion (deduping) while
supporting a distributed and dynamic membership. The long-term persistence
of news messages is evident in ``Google Groups,'' 
a Usenet archive with posts dating from May 1981
to the present ~\cite{google:usenet1}.  Even though
blogs  and forums have supplanted Usenet in recent years, many communities still
actively use moderated news groups for discussion and awareness.
Although email is not usually publicly archivable, it is ubiquitous and frequent.
For example, our departmental SMTP email server alone
averaged over 16,000 daily outbound emails
to more than 4000 unique recipient servers during a 30-day test period.
Given enough time,
attaching repository contents to outbound emails may prove to be an
effective way to disseminate contents to previously unknown locations.
Open source products for news (``INN'') and email (``sendmail'' and ``Postfix'')
are widely installed, so including a preservation function would not impose
a significant additional administrative burden. 

These approaches do not address the more complex aspects of
preservation such as format migration and emulation, but they do provide
alternative methods for refreshing the repository contents to a variety of
recipients, known and unknown.  There may be quicker and more direct methods
of synchronization for some repositories, but the proposed methods have
the advantage of working with firewall-inhibited organizations
and repositories without public, machine-readable interfaces.
For example, many organizations have web servers which are accessible only
through a VPN, yet email and news messages can freely travel between these servers
and other sites without compromising the VPN.
Piggybacking on mature software implementations of these other, widely deployed
Internet protocols may prove to be an easy and potentially more
sustainable approach to preservation.

\section{Related Work}
Digital preservation solutions often require sophisticated system
administrator participation, dedicated archiving personnel,
significant funding outlays, or some combination of these.  Some
approaches, for example Intermemory~\cite{goldberg:intermemory},
Freenet~\cite{clark:freenet1}, and Free Haven~\cite{dingledine:freehaven},
require personal sacrifice for public good in the form of donated
storage space.  However, there is little incentive for users to incur
such near-term costs for the long-term benefit of a larger, anonymous
group.  In contrast, LOCKSS ~\cite{maniatis:lockss} provides
a collection of cooperative, deliberately slow-moving caches operated
by participating libraries and publishers to provide an electronic
``inter-library loan'' for any participant that loses files.  Because it
is designed to service the publisher-library relationship, it assumes
a level of at least initial out-of-band coordination between the
parties involved.  Its main technical disadvantage is that the protocol
is not resilient to changing storage infrastructures.  The rsync
program~\cite{tridgell:rsync} has been used to coordinate the contents of
digital library mirrors such as the arXiv eprint server but
it is based on file system semantics and cannot easily be abstracted to
other storage systems.  Peer-to-peer services have been studied as a basis
for the creation of an archiving cooperative among digital repositories
~\cite{cooper:peertrading}.  The concept is
promising but their simulations indicated scalability is problematic
for this model.  The Usenet implementation \cite{eternity:phrack} of the
Eternity Service \cite{eternity:pragocrypt} is the closest to the methods
we propose.  However, the Eternity Service focuses on non-censorable
anonymous publishing, not preservation per se.

\section{The Prototype Environment}
\begin{figure}
	\centering
	\includegraphics[width=6in]{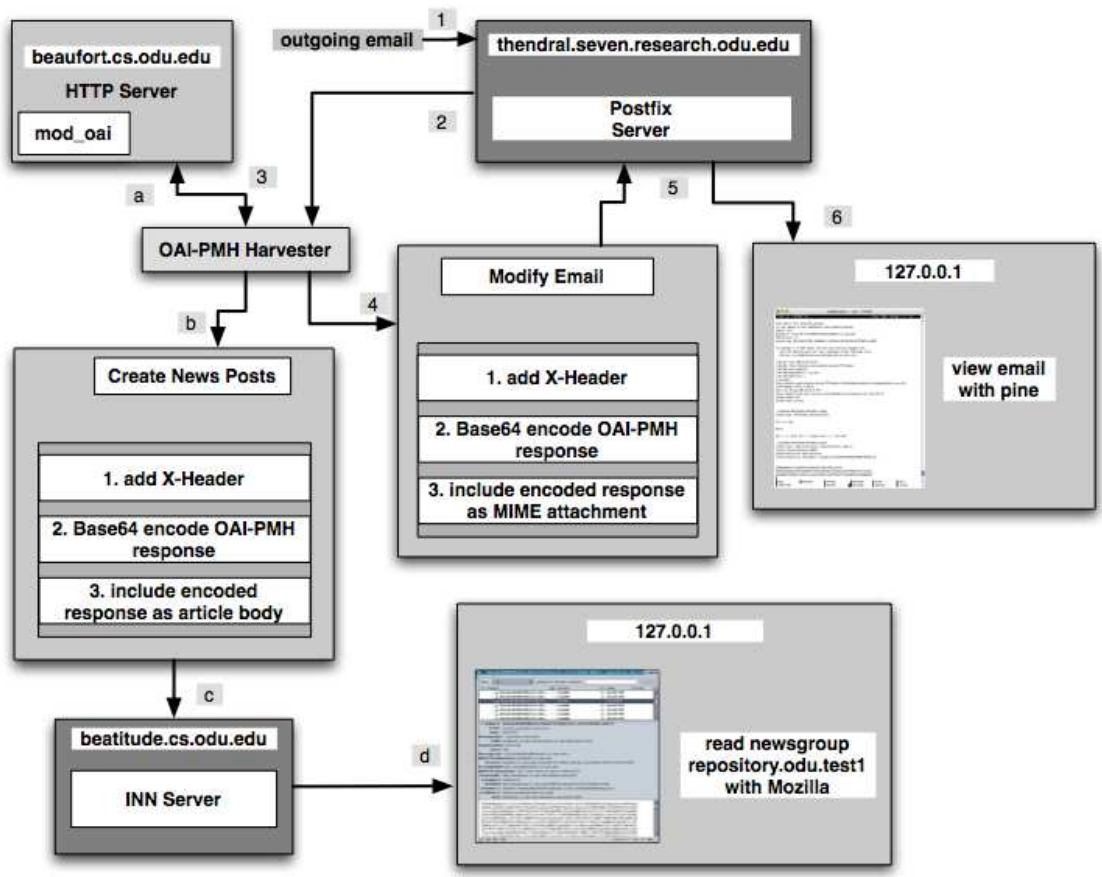}
	\caption{The prototype environment}
	\label{archpic}
\end{figure}

We began by creating and instrumenting a prototype system using
popular, open source products: Fedora Core (Red Hat Linux) operating system;
an NNTP news server (INN version 2.3.5); 
two SMTP email servers, Postfix (version 2.1.5) and sendmail (version 8.13.1);
and
an Apache web server (version 2.0.49)
with the mod\_oai module installed \cite{nelson:modoai}.
Figure \ref{archpic} illustrates the prototype environment we installed. 
No server was dedicated to news or
mail; they also provided services to other users, including project
development environments, operational software, and web services.
\emph{mod\_oai} is an Apache module that provides Open Archives Protocol for
Metadata Harvesting (OAI-PMH) \cite{lagoze:oaipmh} access to a web server.
Unlike most OAI-PMH implementations, mod\_oai does not just provide
metadata about resources, it can encode the entire web resource itself
in MPEG-21 Digital Item Declaration Language \cite{bekaert:didl} and
export it through OAI-PMH. 
\afterpage{\clearpage}

There are many kinds of digital libraries and a wide variety of repository file formats in use today. 
Web access to library content is becoming more common, keeping pace with Internet growth
and facilitated by the many tools which convert hitherto proprietary content
to HTML, PDF, or other
web-compatible formats. In keeping with this trend toward Internet accessibility,
we created a small repository of web resources consisting of 72 files
in HTML, PDF and various image (GIF, JPEG, and PNG) formats.  
We used our own synthetic web site creation tool, building small HTML pages
containing a table, some random text, and a few images as well as links to other
pages in the web site. The PDF files were simple text pages.
The files were
organized into a few subdirectories with file sizes ranging from less than
1 kilobyte up to 1.5 megabytes, and the total web site size was approximately 30 MB.

For the NNTP part of the experiment, we configured the INN news server with common
default parameters: messages could be text or binary;
maximum message life was 14 days; and direct news posting was allowed.
For email, we did not impose restrictions on the size of outgoing attachments
and messages. We created the archive messages within the Postfix 
environment and sent/received the messages using sendmail.
Using custom NNTP and SMTP tools written mainly in Perl and which were operated
from remote clients, we harvested the entire repository over 
100 times with each tool.

We took advantage of OAI-PMH and the flexibility of email and news to
embed the URL of each record as an X-Header within each message.
X-Headers are searchable and human-readable, so their contents give a
clue to the reader about the purpose and origin of the message. Since
we encoded the resource itself in base64, this small detail can be
helpful in a forensic context. If the URL still exists, then the X-Headers
could be used to re-discover the original resource.
Table ~\ref{xhead} is a set of
actual X-Headers added to an archival message,
to facilitate discovery
and recovery of the replicated record.
\begin{table}
\begin{center}
	\caption{Example of human-readable X-headers added to archival messages}
	\begin{tt}
	\begin{tabular}{ l }
		\\
		\hline
	 X-Harvest\_Time: 2006-2-15T18:34:51Z  \\
	 X-baseURL: http://beatitude.cs.odu.edu:8080/modoai/ \\ 
	 X-OAI-PMH\_verb: GetRecord \\
	 X-OAI-PMH\_metadataPrefix: oai\_didl \\
	 X-OAI-PMH\_Identifier: http://beatitude.cs.odu.edu:8080/1000/pg1000-1.pdf\\
	 X-sourceURL: http://beatitude.cs.odu.edu:8080/modoai/?verb=GetRecord \\
	 \&identifier=http://beatitude.cs.odu.edu:8080/1000/pg1000-1.pdf \\
	 \&metadataPrefix=oai\_didl \\
	 X-HTTP-Header: HTTP/1.1 200 OK \\
		\hline 
	\end{tabular}
	\end{tt}
	\label{xhead}
\end{center}
\end{table}
Both the NNTP and SMTP repository harvesting methods use the following algorithm: 
\begin{center}
\begin{tabular}{|l l l l| }
	\hline
	& & &  \\
	& \multicolumn {2}{l}{\texttt{for $r=1$ \textsc{to r}}}& \\
	& & \texttt{read repository record $r$}& \\
	& & \texttt{format $r$ (mail or news)}& \\
	& & \texttt{$r = $ base64($r$)}& \\
	& & \texttt{$r = r +$ X-headers}& \\
	& & \texttt{transmit $r$}& \\
	& \multicolumn {2}{l}{\texttt{end-for}}& \\
	& & &  \\
	\hline
\end{tabular}\label{steps}
\end{center}	
Figure~\ref{archpic} graphically illustrates the process for each replication method.
In sections \ref{newsProto} and \ref{emailProto} we discuss 
the specific details of, and differences between, using
news and email for repository replication.

\subsection{The News Prototype}\label{newsProto}
A testament to Internet diversity, 
Usenet groups exist in many formats. For our experiment,
we created a \emph{moderated} newsgroup which means that
postings must be authorized by the newsgroup owner. This is
one way newsgroups keep spam from proliferating on the
news servers. We also restricted posts to selected IP addresses
and users, further reducing the ``spam window'' 
and ensuring our live experiment would not be compromised by
external news agents and Usenet enthusiasts. Since the news server was
running on a live system used by many people not participating in
the project, controlling access was important. 
For the experiment,
we named our newsgroup ``repository.odu.test1,''
but groups can have any naming scheme that makes sense to
the members. For example, a DNS-based scheme that used
``repository.edu.cornell.cs'' or ``repository.uk.ac.soton.psy''
would be a reasonable naming convention. 

Using the algorithm outlined \vpageref[above]{steps}, 
we created a news message for each record in the repository 
(Cf. Appendix).  We also collected statistics on (a) original record size vs.
posted news message size; (b) time to harvest, convert and post a message;
and (c) the impact of line length limits in news posts.
Our experiment showed high reliability for replicating
using NNTP. 100\% of the records arrived intact on the target news server, ``beatitude.'' 
In addition, 100\% of the records were almost instantaneously mirrored on
a subscribing news server (``beaufort''). A network outage during one
of the experiments temporarily prevented communication between the two news servers,
but the records were replicated as soon as connectivity was restored.
Retrieving messages was as simple as pointing a news reader to the news server,
and subscribing to the ``repository.odu.test1'' news group. 

\subsection{The Email Prototype}\label{emailProto}
\begin{figure}
  \centering
  \subfigure[Outbound mail]{\label{fig:outbound-a}\includegraphics[totalheight=2in]{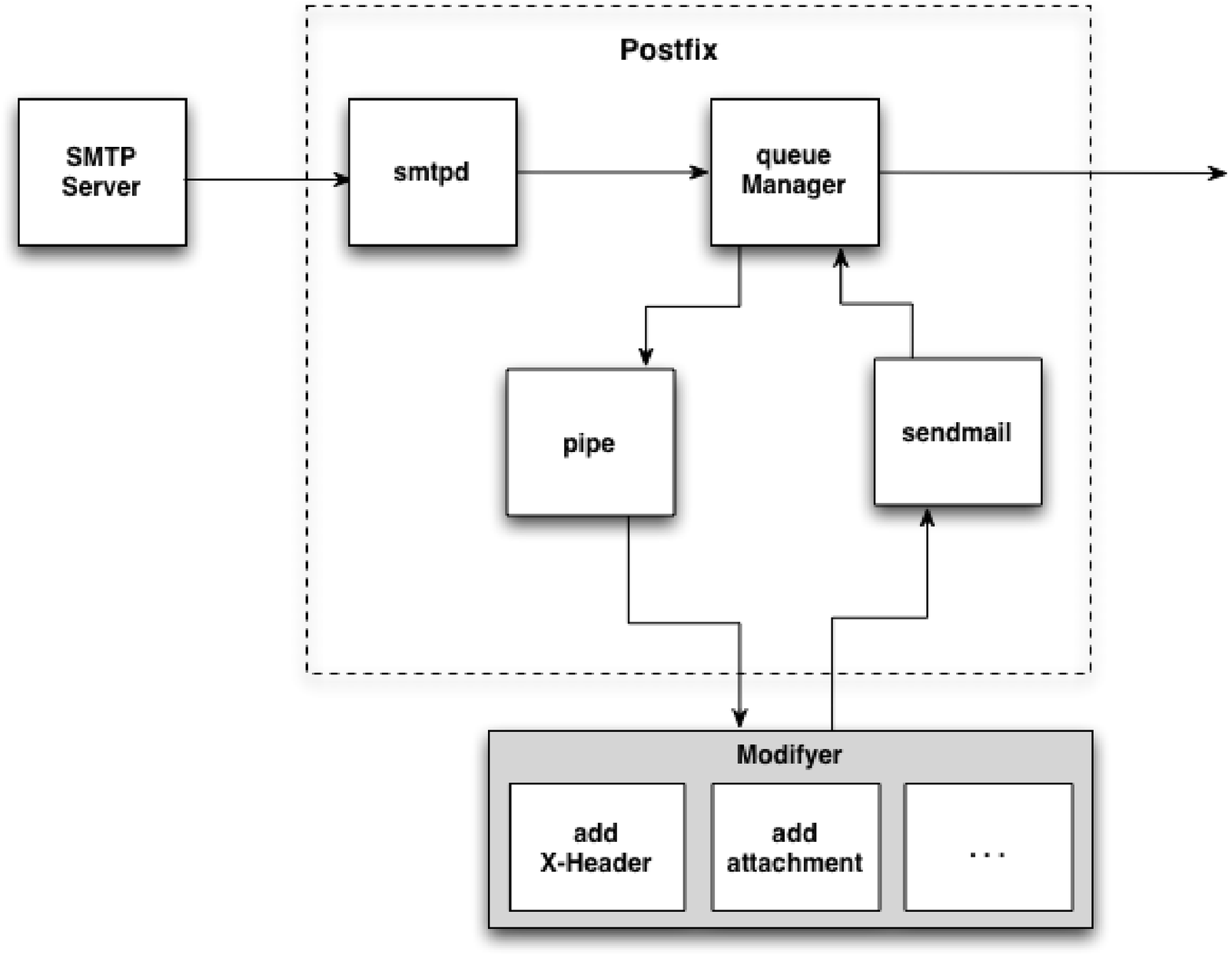}}
  \subfigure[Inbound mail]{\label{fig:inbound-b}\includegraphics[totalheight=2in]{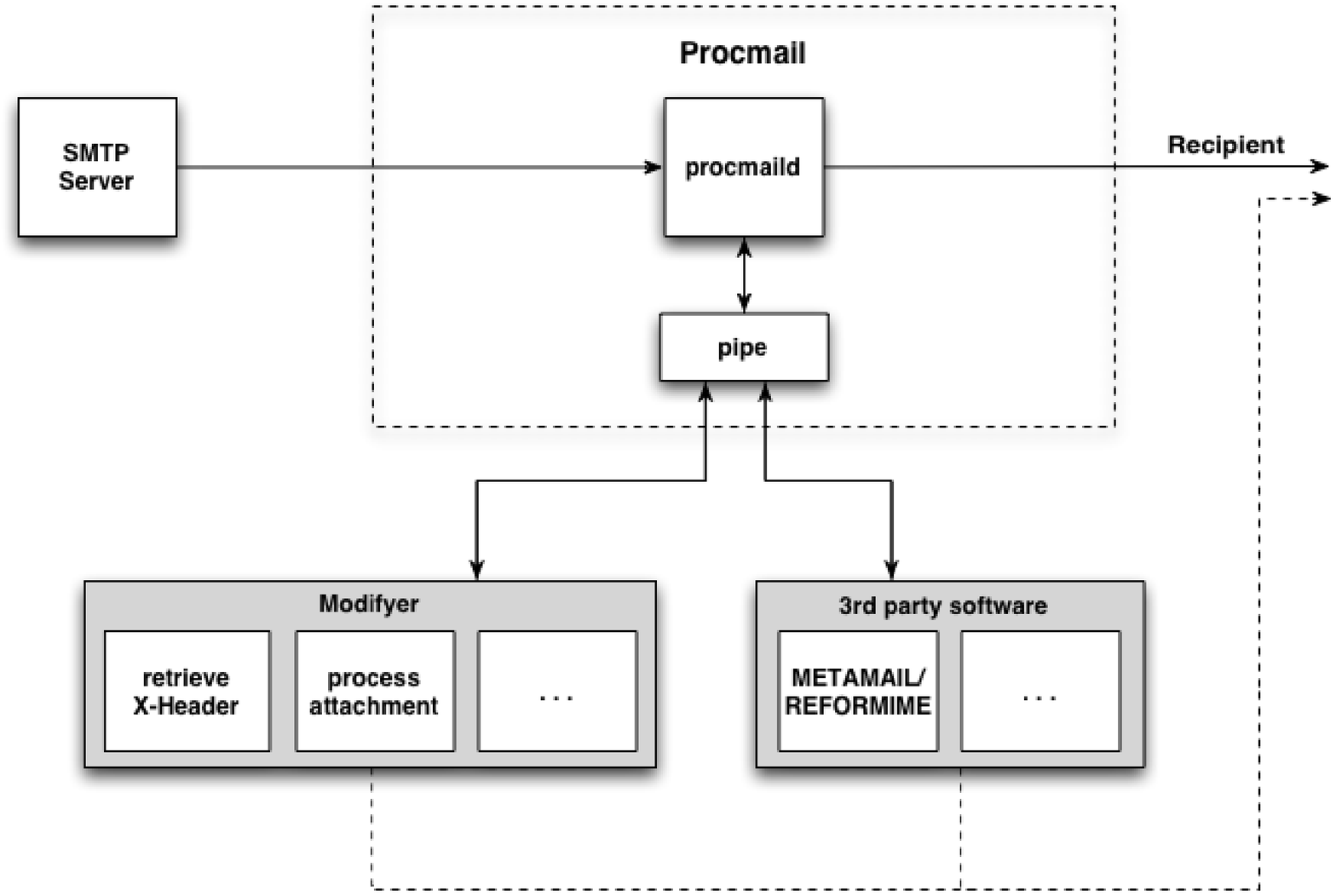}}
  \caption{Archiving using SMTP}
  \label{fig:smtpTool}
\end{figure}

The mechanics of taking an email message
from the email queue,
attaching the archive content, and reinserting it into the queue are depicted
in Figure~\ref{fig:outbound-a}. The corresponding extraction of the archive
attachment can be seen in Figure~\ref{fig:inbound-b}.
We ran live tests, using Postfix
mail servers and a test archive to gather our data. Note the OAI-PMH style
X-headers that are a part of the email message; these are similar
to the X-headers of the news-method messages. The few differences are due
to the specific header limitations and requirements of each protocol. 

\begin{figure}
        \centering
		\includegraphics[totalheight=2.5in]{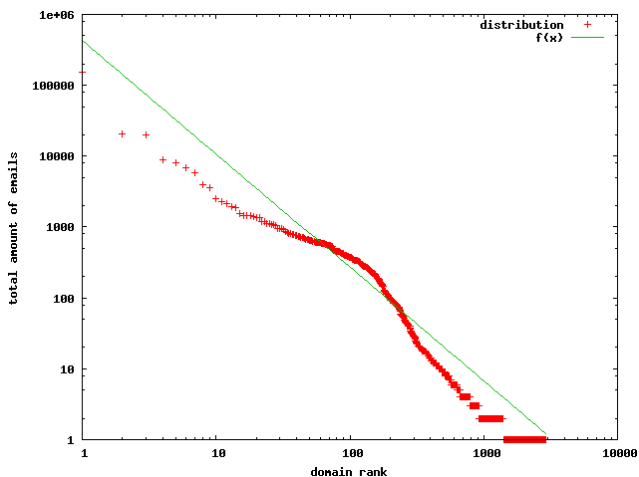}
		\caption{Email distribution by domain follows a power law}
        \label{fig:emaildis}
\end{figure}

Archiving records by piggybacking on normal email traffic requires
sufficient volume to support the effort.  Analysis of outbound email
traffic from our department during a 30-day period showed 505,987 outgoing
messages to 4,493 unique hosts, with a daily mean frequency of 16,866
emails and a standard deviation of 5,147.  In Figure~\ref{fig:emaildis},
the total number of emails sent to each domain is shown, along with a
curve fit.  A typical power law relationship was evident between the
domain's rank and email volume sent to that domain.

\begin{equation}
	V_\kappa=c(\kappa^{-b})
	\label{equ:powerlaw1}
\end{equation}

Using the curve fit shown in Figure~\ref{fig:emaildis}, $b=1.6$.
Please see the Appendix for the list of the top 50 domains and volume
of email sent to each.  For further discussion it becomes necessary
to calculate the amount of emails that are actually sent to a certain
domain per day. The Euler zeta function:

\begin{equation}
\zeta(b)=\displaystyle\sum_{n=1}^\infty\frac{1}{n^b}
\label{1Euler}
\end{equation}
can be used to derive the constant $c$ regarding the overall email traffic
volume $V$:
\begin{equation}
	\displaystyle c = \frac{V}{\zeta(b)}
\label{equ:1cval}
\end{equation}

\begin{figure}
	\centering
	\includegraphics[totalheight=2.5in]{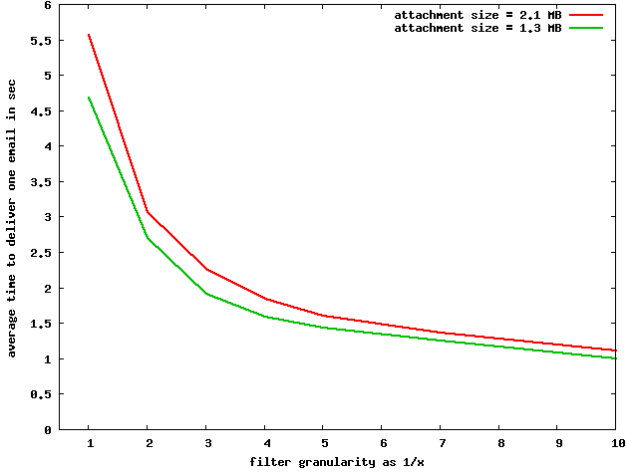}
	\caption{Effect of granularity on average delivery time of one email}
	\label{fig:gran}
\end{figure}

There are a number of processing parameters which can be tweaked while running
the prototype. One factor is what we call ``granularity'' ($G$ in Table~\ref{simVars}).
This factor is in our prototype by definition always unequal to zero. The "normal"
case would be $G=1$ which means every single email is selected to get an attachment
of an harvested object. G can be less than zero in which case only every $G^{th}$
email is attached with an replication object. If, for example, $G=0.5$ only
every other email is selected. On the other hand G can be greater than one, e.g.
$G=3$ in which case three objects would be attached to every single email. Granularity
$G$ consequently either functions as a damping or accelerating factor considering the 
pace of repository replication. The effect of $G < 1$ on the average time to deliver one
email is shown in Figure~\ref{fig:gran}. With a lower G (less emails selected
for an attachment) the average delivery time decreases.

The prototype is further able to maintain a history list (pointer) for each destination
site. Once this feature enabled, it guarantees that one destination domain does not
receive duplicate records. The concept of a history pointer is further explained in
section \ref{emailDeriv}.

\subsection{Prototype Results}
Having created tools for harvesting the records from our sample digital library, 
and having used them to replicate the repository, 
we were able to measure the results.
How fast is each prototype and what penalties are incurred?

Using our NNTP prototype replication tool, we tested
posting messages in a variety of sizes. The live experiment
was run more than 20 times during a course of 6 months.
The total time ($T_{news}$)
to harvest a record, encode it in base64, transmit it, and post it
to the news server 
ranged from 0.5 seconds (12 \textsc{kb}) to 26.4 seconds (4.9MB).
Of course, the total time to complete a baseline harvest
of the repository varied
with the bandwidth available during each experiment,
ranging from 22.7 minutes to 30.9 minutes  
with a mean time of 23.8 minutes, standard deviation of 2.6
minutes, and median time of 22.9 minutes.

\begin{figure}
        \centering
		\includegraphics[totalheight=2.5in]{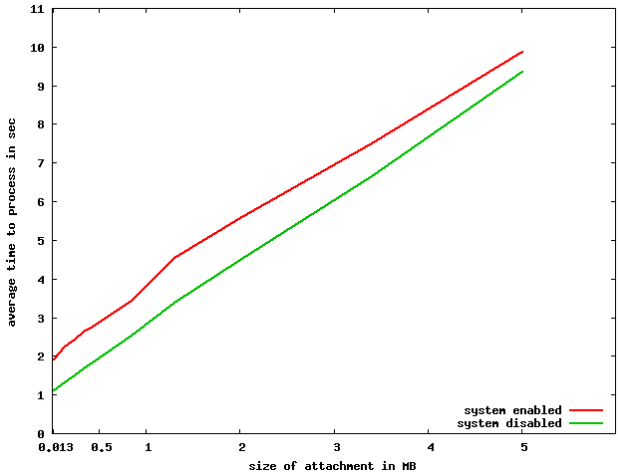}
		\caption{Penalty on average delivery time of one email}
        \label{fig:filterpen}
\end{figure}

In our email experiment, we measured approximately a 1 second delay in
processing email attachments of sizes up to 5MB (see Figure~\ref{fig:filterpen}). 
Since the repository consisted of only 72 files and each file was
less than 5MB $T_{email}$, the time to complete a baseline harvest
using email, is rapid: Only 72 emails need to be generated locally,
which is a small fraction of the normal email traffic 
generated by the department. 

Besides the trivial linear relationship
between repository size and replication time, we found that even
very detailed X-Headers do not add a significant burden to the process.
Not only are they small relative to record size (a few bytes vs. kilobytes or more),
but they are also quickly generated (less than 0.001 seconds per record)
and incorporated into the archival message. Both NNTP and SMTP protocols
are robust, with most products (like INN or sendmail) automatically
handling occasional network outages or temporary unavailability of the
destination host. 
Our experimental results formed the basis of a series of simulations
using email and Usenet to replicate a repository.

\section{Simulating The Archiving Process}
When transitioning from live, instrumented systems to simulations, there are a number of
variables that must be taken into consideration in order to arrive at realistic
figures (Table ~\ref{simVars}). Repositories vary greatly in
size, rate of updates and additions, and number of records. 
Regardless of the archiving method, a repository will have specific policies
(``Sender Policies'') covering the number of copies replicated;
how often each copy is refreshed; whether intermediate updates
are sent between full backups; and other institutional-specific
requirements such as geographic location of archives and ``sleep time'' (delay)
between the end of one completed archive task and the start of another.
The receiving agent will have its own ``Receiver Policies''
such as limits on individual message size, length of time messages
live on the server, and whether messages are processed by batch or 
individually at the time of arrival.
\begin{table}[h!]
	\centering
	\caption{Simulation variables}
    \begin{tabular}{l  l  l}
		\multicolumn{3}{l}{ }\\
	\hline
	\multirow{4}{*}{Repository} & $R$ & Number of records in repository \\
	& $R_{\overline{s}}$ & Mean size of records \\
	& $R_a$ & Number of records added per day \\
	& $R_u$ & Number of records updated per day \\
	\hline
	\multirow{4}{*}{Usenet} & $N_{ttl}$ & News post time-to-live \\
	& $S$ & ``Sleep'' time between baseline harvests \\
	& $Q_{news}$ & Records postable per day via news \\
	& $T_{news}$ & Time in days to complete a baseline using news\\
	& $TR_{news}$ & Total number of records replicated using news\\
	\hline
	\multirow{7}{*}{Email} & $G$ & Granularity (records per email)\\
	& $Q_{email}$ & Records postable per day via email \\
	& $TR_{email}$ & Total number of records replicated using email\\
	& $\kappa$ & Rank of receiving domain\\
	& $c$ & Constant derived from Euler Zeta function\\
	& $b$ & Power law exponent\\
	& $h$ & History pointer \\
	\hline
	& & \\
	\end{tabular}
	\label{simVars}
\end{table}

A key difference between news-based and email-based replication is the active-vs-passive
nature of the two approaches. This difference is reflected in the policies and how
they impact the archiving process under each method. 
A ``baseline,'' refers to making a complete snapshot of
a repository. A ``cyclic baseline'' is the process of repeating the
snapshot over and over again ($S=0$), 
which may result in the receiver
storing more than one copy of the repository. 
Of course, most repositories are not static. Repeating baselines will capture new
additions ($R_a$) and updates ($R_u$) with each new baseline. 
The process could also  ``sleep'' between baselines ($S>0$),
sending only changed content during the interim, or none at all.
In short, the changing nature of the repository can
be accounted for when defining its replication policies.

\subsection{Archiving Using NNTP}
\begin{figure}
	\centering
	\subfigure[Generalized NNTP timeline]{\label{newsTimeline}\includegraphics[totalheight=1.6in]{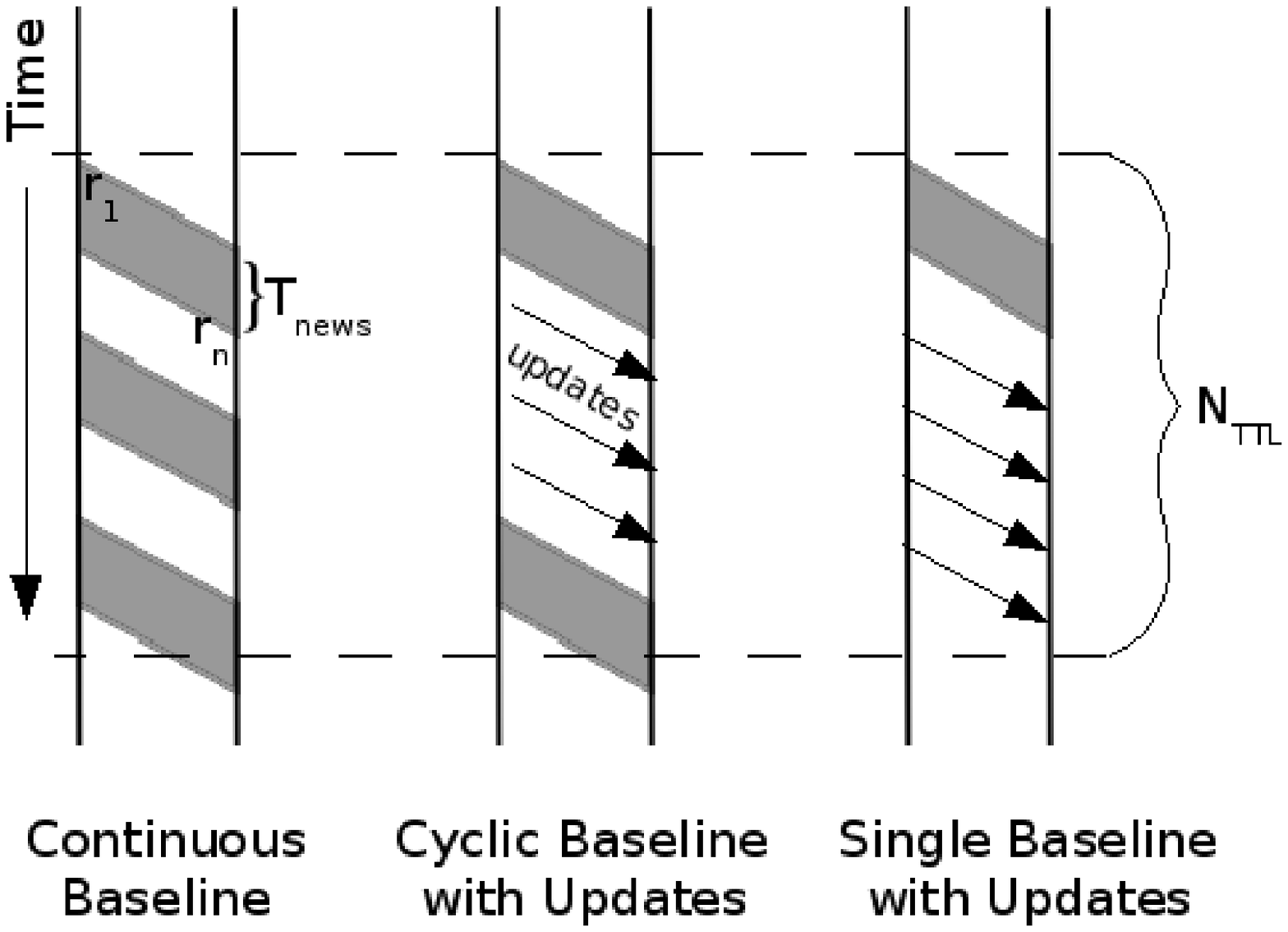}}
	\subfigure[Varying $N_{ttl}$]{\label{varyNttl}\includegraphics[totalheight=1.6in]{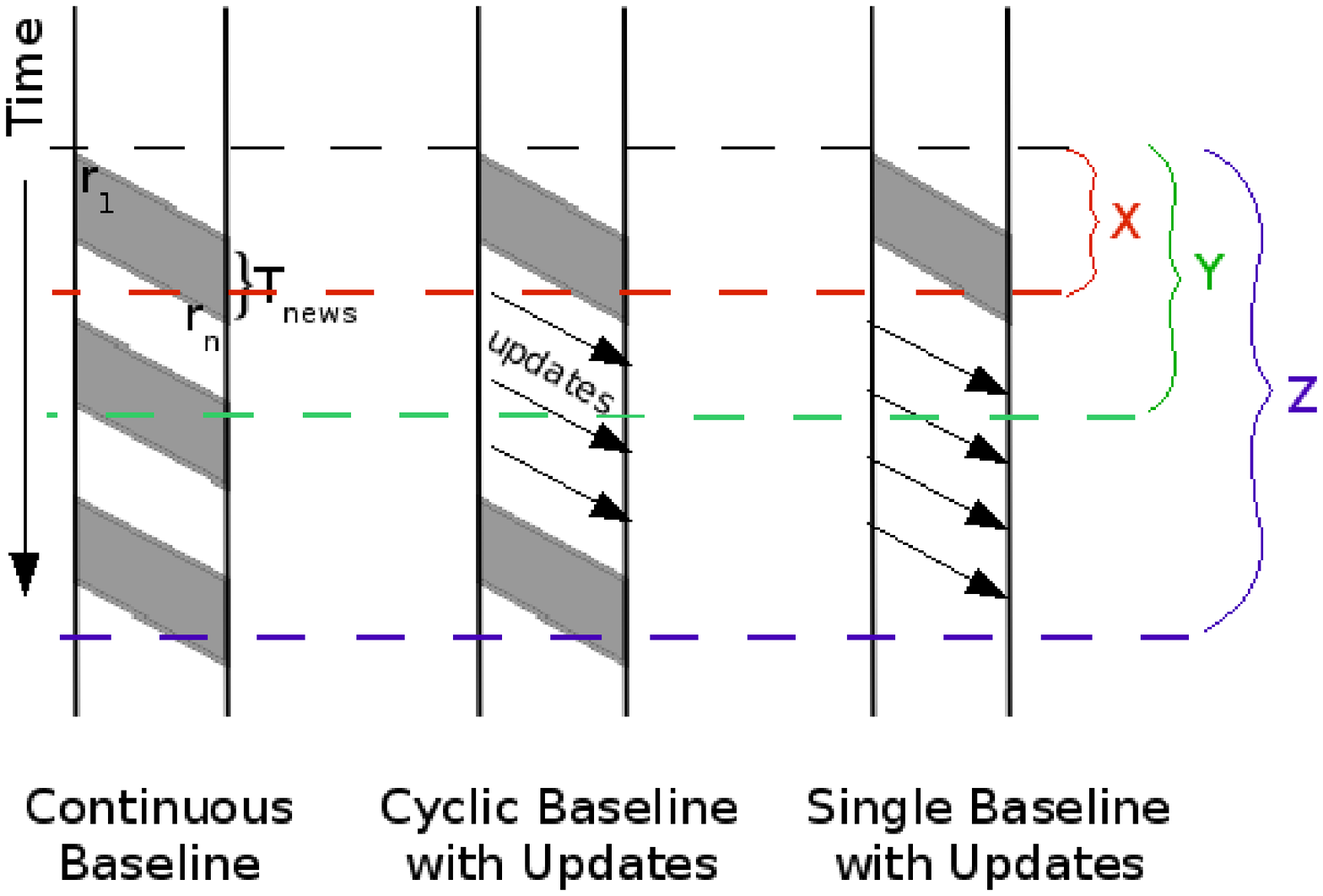}}
	\label{fig:nttl}\caption{Impact of sender \& receiver policies on NNTP replication}
\end{figure}

The time to complete a baseline using news is obviously constrained
not only by its modification rate ($R_a$ and $R_u$), but also
by the size of the repository and the speed of the network.
Consider Figure \ref{newsTimeline} which illustrates the generalized 
replication timeline for three different sender policies. 
Baseline replication is only successful when
the news server message life ($N_{ttl}$) is larger than $T_{news}$.
Figure \ref{varyNttl} shows how different message life limits
can impact the feasibility of archiving the repository on a news
server under different sender policies.
The red line (marked ``X'') shows a message life that is
not long enough for a single baseline to complete - i.e., that
$T_{news}$ is too large for the target news server. Line ``Y''
(the green line) represents a longer message life than line X,
but there is still not enough time for the server to ``sleep'' between
baseline archives. If the harvest restarts immediately on completion
of the first baseline, a full copy can be maintained on the news
server despite its message deletion rate. Repository growth could
quickly outpace this balance. Finally, line ``Z''
(the blue line) is long enough to allow two complete baselines
(copies) to be sent, with
a short sleep period between the baselines. 
A successful NNTP-based
replication strategy will balance $N_{ttl}$, $T_{news}$, and
the repository's modification rate ($R_{a} + R_{u}$).

Working with the variables from Table~\ref{simVars}, we can develop a
general formula to estimate the total number of records harvested from the repository
and posted as news articles during
$D$ days.  These equations capture only discrete values
and not transmissions in progress:
\begin{equation}
	TR_{news} = \displaystyle\left(\sum_{n=1}^{Max K} {W_k}\right)\leq{D}
	\label{equ:recSum}
\end{equation}
\begin{equation}
	W_k = \displaystyle\left(1+\frac{(R_a + R_u)}{Q_{news}}\right)W_{(k-1)}
	\label{equ:wk}
\end{equation}
\begin{equation}
	W_{1} = \frac{R}{Q_{news}} + S 
	\label{equ:w1}
\end{equation}
For the sleep cycle, $S$, the value varies by sender policy:
\begin{equation}
	S\: =\: 0\: \Longrightarrow{\textnormal{continuous baseline}}
	\label{equ:continuous}
\end{equation}
\begin{equation}
	S\: =\: D\: \Longrightarrow{\textnormal{cyclic baseline every $D$ days with updates}}
	\label{equ:cyclic}
\end{equation}
\begin{equation}
	S\: =\: \infty\: \Longrightarrow{\textnormal{single baseline only}}
	\label{equ:baseline}
\end{equation}
The total number of records currently replicated
at a particular news server $N$ on a given
day $D$ takes into account 
the life time ($N_{ttl}$) of news messages on that server:
\begin{equation}
	{TR_{news}\:at\:server\:N} = TR_{news}(D) - TR_{news}(D-N_{ttl})
	\label{equ:tot_rec}
\end{equation}

Nearly all repositories will have daily updates
and new additions that need to be accounted for when determining $T_{news}$. 
Even ``static'' repositories which do not accept new entries are likely to
have a certain amount of periodic record modification as errors, for example,
are found and corrected.
A larger time gap between baseline harvest completion and news message expiration will
give the harvesting repository more ``room to grow'' before the two timelines
collide.

NNTP is an older protocol, with limits on line length and content which
impact building the news messages. Converting binary content to base64 overcomes 
such restrictions but at the cost of increased file size (one-third) and replication time.
Even though storage costs continue to decline, a complete baseline harvest with its
associated metadata and base64 encoding could prove too large for a news server to
support. On the other hand, the web infrastructure has a number of participants
(Google Groups, for example) which are interested in maintaining cached versions of even
very large sites. In this case, a single baseline with updates could prove to be an
acceptable strategy for a repository.

\subsection{Archiving Using SMTP}
One major difference in using email as the replication tool instead of news is that
email is passive, not active: 
the email approach relies on \emph{existing} traffic between the host site and one or
more target destination sites. Fortunately, the prototype is able to attach
files automatically with just a small processing delay penalty of less than 1 second.
As it turns out though, maintaining a replication list (history pointer)
for each destination site is critical if a baseline harvest is to be completed.

Using the variables defined in Table~\ref{simVars},
we can develop a general formula to estimate the total number
of records harvested in $D$ days to a specific destination:
\begin{equation}
	TR_{email} = \displaystyle\sum_{n=1}^{D}Q_{email} \times h({\textsc{d}})
	\label{equ:email}
\end{equation}
\begin{equation}
	Q_{email} =\displaystyle \left(\frac{c}{\kappa^{1.6}}\right)\times G
\end{equation}\label{equ:qmail} 
\begin{equation}
	0 < h({\textsc{d}}) < 1 \Longrightarrow {\textnormal{no record history pointer maintained}} 
\end{equation}\label{equ:h0}
\begin{equation}
	h({\textsc{d}}) = 1 \Longrightarrow {\textnormal{record history maintainted}}
 \end{equation}\label{equ:h1}
If the history list is maintained for every receiver domain, then the pointer value
is equal to 1, as indicated in Equation \ref{equ:h1}; but if the history pointer is \emph{not}
maintained, then the value varies between 0 and 1 (zero and one) as shown in
Equation \ref{equ:h0}.
The value of $h({\textsc{d}})$ is 
derived in Equation \ref{probnewemail}.
Unlike news, which is \emph{time} oriented,
the email approach is \emph{destination} oriented.
Granularity ($G$) and history-pointer values ($h$) are important
factors when calculating the replication estimate.
Completing a baseline using email is subject to the
same constraints as news - repository size, number of records, etc. -
but is particularly sensitive to changes in email volume. For example,
holidays are often used for administrative tasks since they are typically
``slow'' periods, but there is little email generated during holidays
so repository replication would be slowed rather than accelerated.
On the other hand, the large number of unique destination hosts means that email is
well adapted to repository discovery through advertising.
In a single day, information about the repository can be disseminated
to thousands of potential preservation partners.

\section{Simulation Results}
In addition to an instrumented prototype, we simulated a repository
profile similar to some of the largest publicly harvestable
OAI-PMH repositories. The simulation
assumed a 100 gigabyte repository with 100,000 items
($R=100000$, $R_{\overline{s}}~=1MB$); a low-end
bandwidth of 1.5 megabits per second;
an average daily update rate of 0.4\% ($R_{u}=400$);
an average daily new-content rate of 0.1\% ($R_{a}=100$);
and a news-server posting life ($N_{ttl}$) of 30 days.
For simulating email replication, our estimates were based
on the results of our email experiments: Granularity $G=1$, 
an average of 16866 total outgoing emails per day,
and the power-law factor applied to the ranks of receiving hosts.
We ran the NNTP and SMTP simulations for the equivalent of 
2000 days (5.5 years).

\subsection{Policy Impact on NNTP-Based Archiving}
News-based replication is constrained primarily by network
capacity and limits imposed by the receiving news server. 
Except for inter-party agreements 
or some other trans-organizational coordination,
the receiver's policies, even when they are known,
are usually unconfigurable by the sender.
A local news server can influence remote servers
by establishing its own $N_{ttl}$, size limits, content type, etc.
A news server may adopt some of the policies
of the source server it is replicating, allowing posts to expire \emph{earlier} than
the local server's $N_{ttl}$, but usually not allowing the posts to live \emph{longer}.
Ultimately, the archivist must consider the balance 
between the repository's capacity to replicate via
NNTP and the news server's ability to support replication.

As Figure~\ref{fig:nttl} illustrated, successful replication depends
on $T_{news}$ being smaller than $N_{ttl}$.
We can estimate $T_{news}$ using the
average record size in the repository ($R_{\overline{s}}$) times the total number of records ($R$)
and the base64 encapsulation factor ($\frac{4}{3}$), divided by the net available bandwidth ($\nu$):

\begin{equation}
	T_{news} = \frac{R\times R_{\overline{s}} \times \frac{4}{3}}{\nu}
	\label{equ:T_news}
\end{equation}
If the lifetime of a posting is shorter than the baseline harvesting time
of the repository ($N_{ttl} < T_{news}$), then that news server will never 
hold a complete copy of the repository on any given day.

Another potential issue is that
the sheer size of the repository may make full-content
replication to a news server impractical because of limits in 
available processing time or host storage capacity, for example.
In such a situation the repository could adopt a ``By-Reference'' archiving policy. This
approach is fast and efficient, since it stores only the metadata for each
repository record rather than the content of the record. 
Using Equation \ref{equ:T_news}, we see that a repository with $R=500,000$
records and per-record metadata of 1 Kilobyte
can be archived in less than 1 day (ignoring updates and additions)
at speeds as slow as a dial-up modem (0.125 Mbps):
\begin{equation}
	0.37 Days = \frac{500,000 Records\times1 \textsc{kb}}{0.125\textsc{Mbps}}
\end{equation}
For very large and/or very active repositories, this kind of ``advertising''
may be the optimum solution.

In general, the probability of a given repository record being currently replicated on a
specific news server $N$ on day $D$ is a function of the number of records posted each day to
the news server ($Q_{news}$), the growth of the repository ($R_a$) during those
$D$ days, and the lifetime
of the record on the server ($N_{ttl}$):
\begin{equation}
		P(r) = \displaystyle\frac{(Q_{news} \times D)- Q_{news}\times(D-N_{ttl})}{R +(D\times R_{a})}
	\end{equation}

\begin{figure}
  \centering
  \subfigure[The first 200 days]{\label{fig:active200}\includegraphics[totalheight=3.5in]{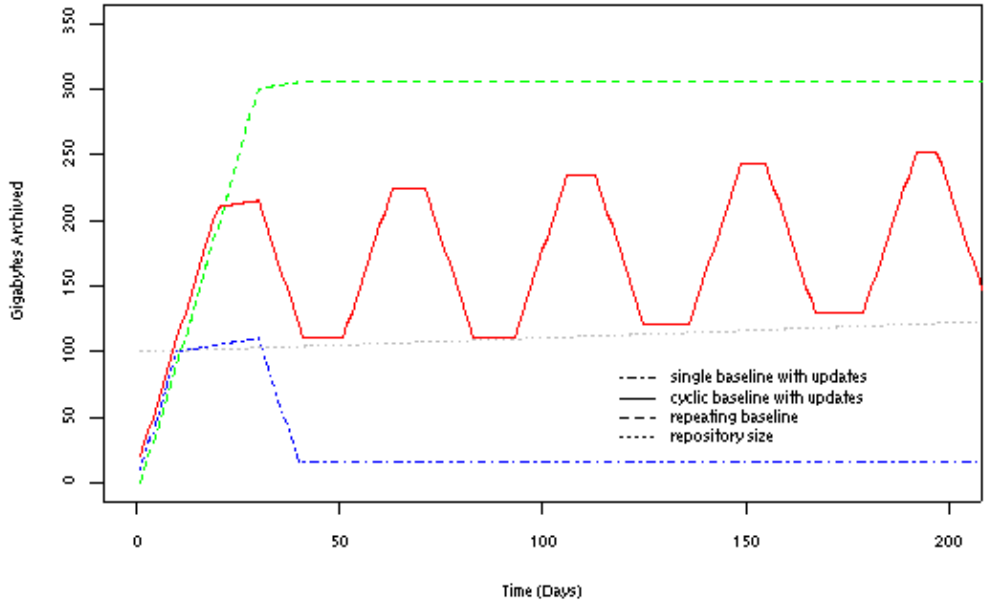}}
  \subfigure[The first 2000 days]{\label{fig:newsPolicies}\includegraphics[totalheight=3.5in]{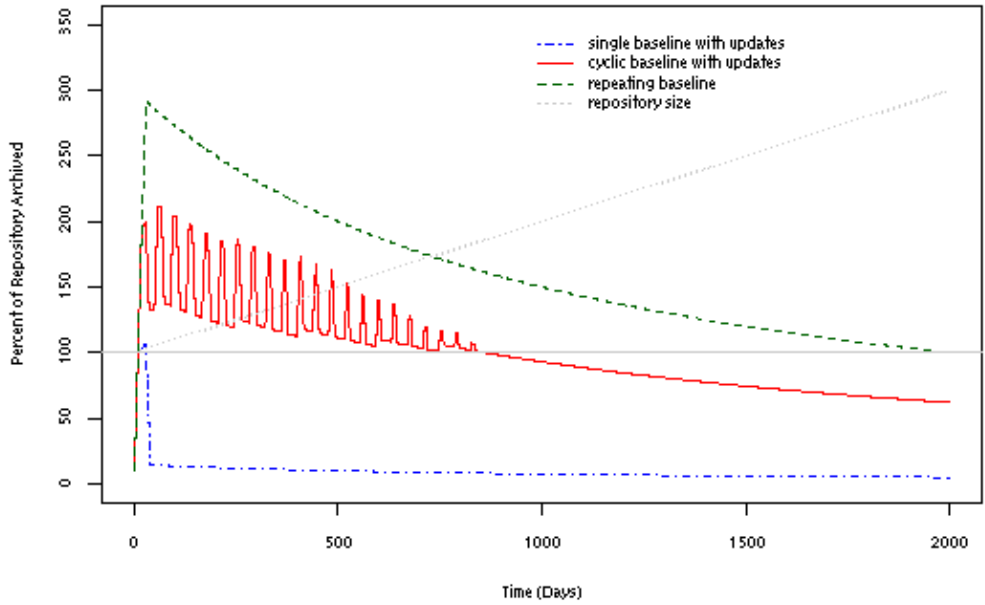}}
  \caption{Replicating an extremely active repository}
  \label{fig:activeRepo}
  \centering
\end{figure}

Figure \ref{fig:newsPolicies} illustrates
how a sufficient grace period ($N_{ttl}$ = 30) 
can support different repository replication (sender) policies.
In one scenario, continuous baselines are transmitted. New and/or modified
records are queued as they occur. 
Both the ``Cyclic Baseline with Updates'' and the ``Repeating Baseline'' approaches
eventually result in a steady-state amount of data
existing on the news server. This amount is approximately equal to the bandwidth
available between the repository and the news server,
and is a gradually declining percentage of the repository as it continues to
grow and modify records. 

For the ``Cyclic Baseline with Updates'' line
in Figure \ref{fig:newsPolicies}, we simulated
a 6-week repeating cycle with certain ``sender policies'':
The entire repository is replicated twice, followed by updates only, then the
cycle is repeated. 
With this approach, the news server maintains between one and 2 full copies of the
repository, at least for the first few years.

The worst replication performance can be seen in the
``single baseline with updates''
line of Figure~\ref{fig:newsPolicies}.
In this third approach, the policy is to make a single baseline copy
i.e., a one-time event, which is followed by only 
record updates and new additions. Even these are eventually
removed from the news server as they reach the limit of $N_{ttl}$.
The result is a rapidly decreasing percentage of repository replication
over time. Eventually, only 30-days' worth of new data exists
on the server, since $N_{ttl}=30$. Usually, this would be a very
small portion of the repository compared with the other two policies,
which can maintain up to $N_{ttl}\times Q_{news}$ versus 
$N_{ttl} \times (R_a + R_u)$, for example.

It is obvious that as a repository grows and other factors such as bandwidth and
news posting time remain constant, the news server eventually contains less than
100\% of the library's content, even with a policy of continuous updates.
Nonetheless, a significant portion of the repository remains replicated for many
years if the news server has a sufficient $N_{ttl}$.

\subsection{Policy Impact on SMTP-Based Archiving}\label{emailDeriv}

SMTP-based replication is constrained not only by the frequency
of outbound emails, but also by the policies adopted by the
repository. Consider the following two sender policies: The first policy
maintains just one queue where items of the repository are being attached
to every $G^{th}$ email regardless of the receiver domain.  This policy
also randomly assigns a record, without maintaining a history pointer of
records which have already been replicated.  This is the easiest policy
to implement since no history pointers are maintained, but it will
take much longer for a particular domain to receive all records since
many duplicate records will likely be sent while unsent record remain.
In the second policy, we have more than one queue where we keep a pointer
for every receiver domain and attach items to every $G^{th}$ email going
out to these particular domains. Thus, domain X will receive a new record
in each attachment. Duplicates will only begin once a baseline to that
domain has completed.  The second policy allows each receiving domain to
converge on 100\% coverage much faster.  However, this efficiency comes
at the expense of the sending repository tracking separate queues for
each receiving domain.

\begin{figure}
        \centering
		\subfigure[Without record history]{\label{fig:cvrg-a}\includegraphics[totalheight=2.25in]{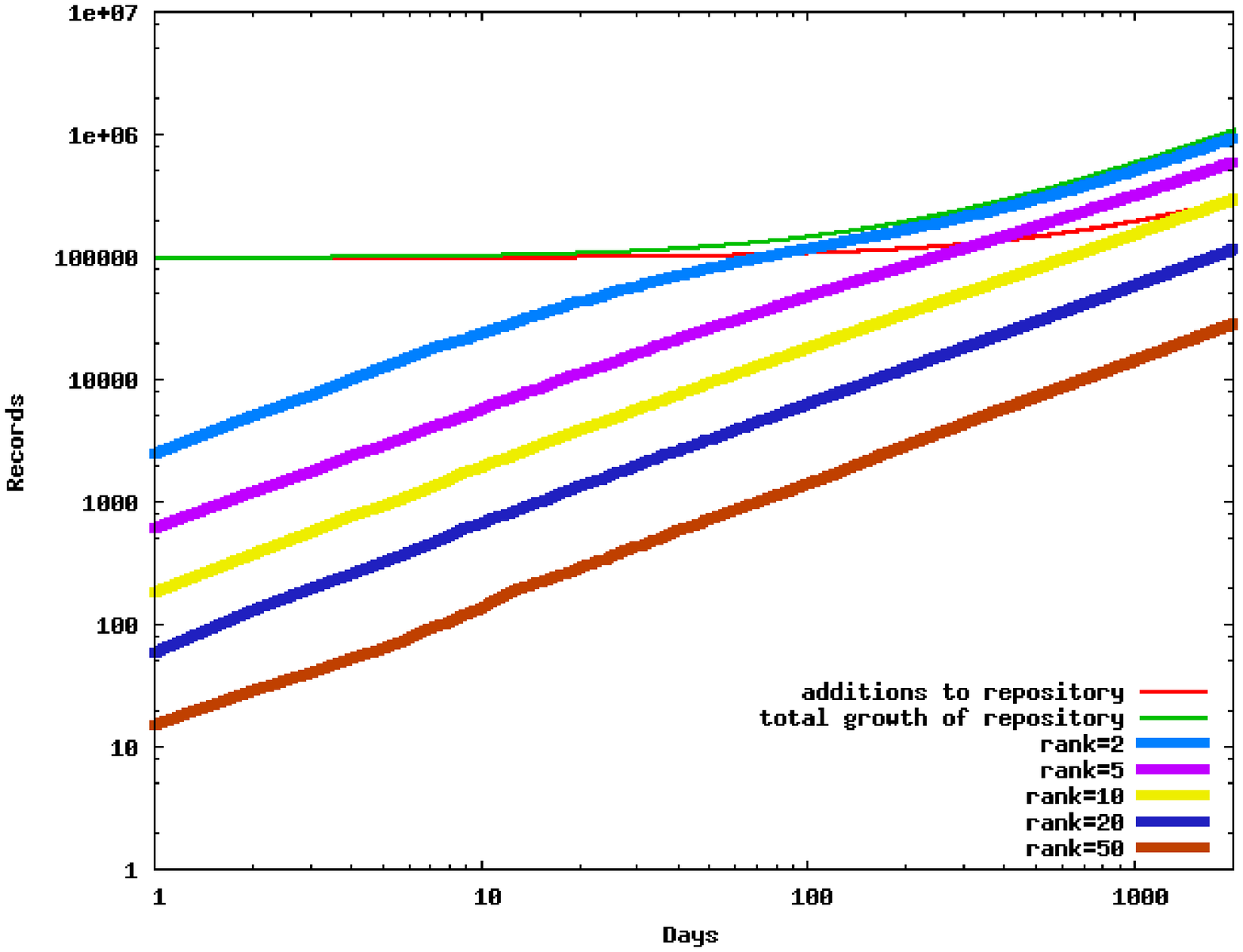}}
		\subfigure[With record history]{\label{fig:cvrg-b}\includegraphics[totalheight=2.25in]{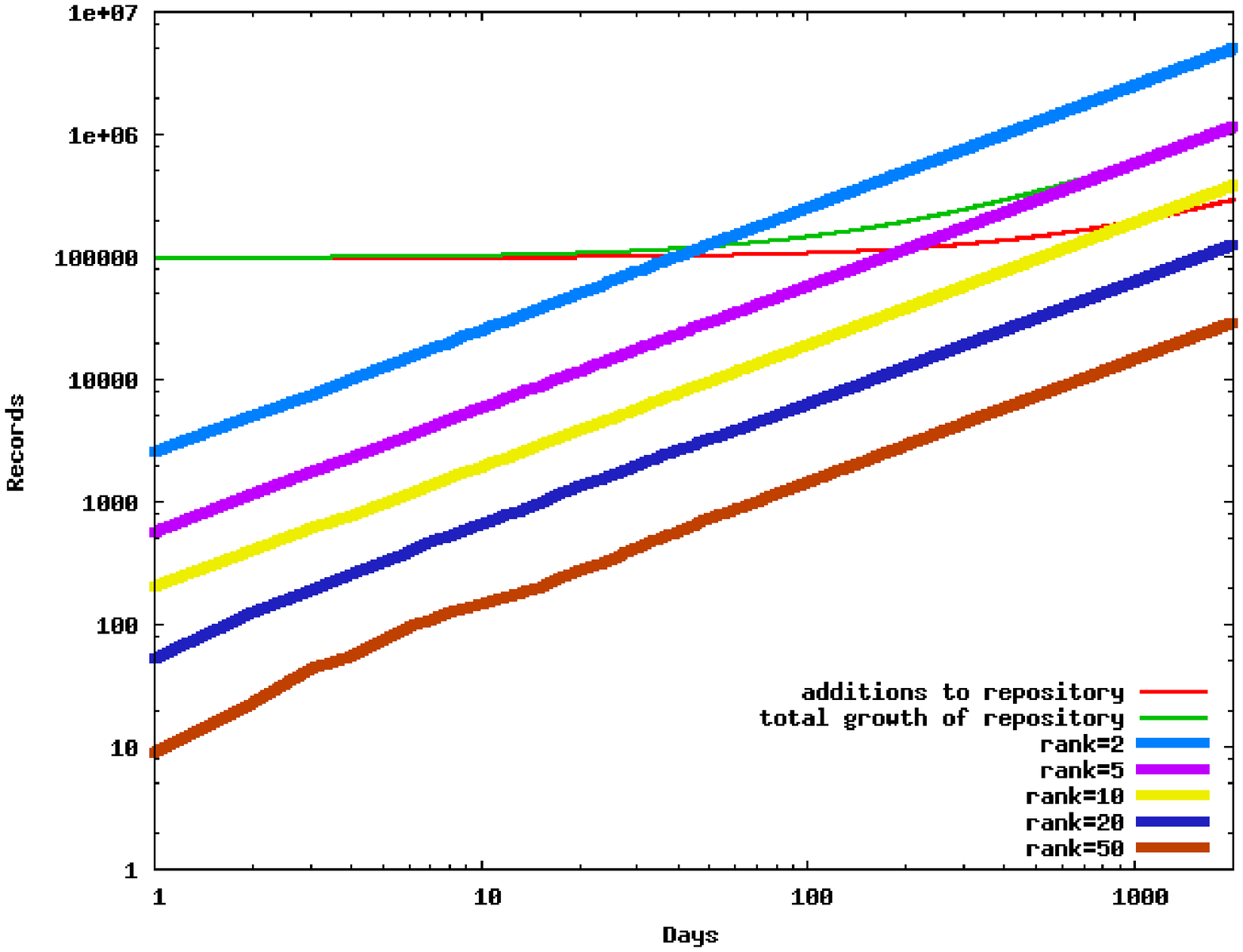}}
        \caption{Time to receive 100\% repository coverage by email domain rank}
        \label{coverage}
\end{figure}

The impact of email's power law distribution is readily seen when
comparing the coverage of higher-frequency ranks (1 through 5, for example)
with lower-frequency ranks. Receiver domains ranked 2 and 3
achieve 100\% repository coverage fairly soon but Rank 20 takes significantly longer
(2000 days with a history pointer), reaching only 60\% if no history pointer is maintained.
Figure \ref{fig:cvrg-a} shows the time it takes
for a domain to receive all files of a repository without the history pointer
and Figure \ref{fig:cvrg-b} shows the same setup with a history pointer.
In both graphs, the $1^{st}$ ranked receiver domain is left out 
because it represents internal email traffic.
%See Appendix \ref{app:email} for a list of the top 50 receiver domains.

Figure \ref{fig:cvrg-a} clearly shows the impact
of failing to maintain a record history.
Since there is a decreasing statistical likelihood of a new record
being selected from the remaining records as the process progresses,
it becomes less and less likely that a baseline harvest can be reached.
Thus, feasibility of replication via email $Q_{email}$ 
is a function
of the receiver's rank ($\kappa$), the granularity ($G$),
and probability based on use of a history pointer ($h$).
Working with the values obtained from our experiments 
where $b=1.6$ and total email volume per day = 16866,
and Equation~\ref{equ:1cval}, we find that the value of the constant
$c$ is 7378.7 ; this value can now be used to determine
the number of emails sent per day for each receiver domain
by rank $\kappa$:
\begin{equation}
	Q_{email}=\frac{7378}{\kappa^{1.6}}\times G
	\label{equ:qEmail}
\end{equation}
A rank of 3, for example, would mean 1,272 emails per day to that host.
The total number of records replicated on day $D$ is:
\begin{equation}
	\displaystyle\sum_{n=1}^{D}Q_{email}\times h(D)
	\label{equ:2probEmail}
\end{equation}

To give us a good opportunity to complete a baseline, we
can set $h = 1$ and $G = 1$.
In other words, we maintain a history pointer, and we do not skip any emails.
This ensures that we do not send duplicate records before a baseline
of the entire repository has been completed, and that we take full
advantage of email traffic to that domain. It is obvious that increasing
G would shift the graphs of both Figures \ref{fig:cvrg-a} and
\ref{fig:cvrg-b} up and decreasing it would just shift them down.
Using these values, we can calculate the probability
that a record has been replicated via email:
\begin{equation}
	P(r) = \displaystyle\frac{(Q_{email} \times D)}{R +(D\times R_{a})}
\end{equation}

What if no history pointer is maintained? In that case, we need to include
the probability that a new record is attached to a given email, meaning
$h(D)$ is no longer one.
The equation for $h$ is is a recursive calculation since it needs to account
for the number of records already sent compared with the number of remaining,
unsent records i.e., non-duplicates. For simplicity, we assume that no
duplicates are sent on the first day (Equation \ref{equ:h_D1}).
\begin{equation}
	h(D)=\displaystyle \frac{[R+((R_u+R_a)\times D)]-Q_{email}}{[R+((R_u+R_a)\times D)]}\times h(D-1)
\label{probnewemail}
\end{equation}

\begin{equation}
	 h(1) = 1 
	\label{equ:h_D1}
\end{equation}

In summary, one can argue that email may not be a practical solution
for repository replication since the lower ranked domains will not get a full
replication of a good sized repository in a reasonable time. The email
approach does have a unique advantage: it offers a large number of hosts
where the repository can be advertised. 

\section{Other Repository Scenarios}

The scenarios we have described so far in this paper involve an unusually
active repository, one which experiences a high rate of change in the
form of new additions and updates to existing records.  Our hypothetical
repository doubles in size in only 1000 days (just under 3 years).
We also used a relatively slow average network speed (most institutions
and even home users will have much higher average bandwidth), and further
added an average 25\% daily network down time. In other words, we stacked
the deck against the NNTP and SMTP replication methods. Despite these
obstacles, the repository continues to be fully replicated on the news
server for over 2 years.

Email as a replication tool poses several problems such as the
passive nature of the process (waiting for emails to be generated),
and uncertainty about the persistence of the record on the
receiving host. On the other hand, the large number of domains
that receive emails make this approach very compatible with
a strategy of preservation-by-advertising: The greater the
number of sites that are aware of a repository, the greater
the likelihood that the repository will be found by interested
users and - perhaps - replicated.

How would these approaches work with other repository scenarios?
If the archive were substantially smaller (10,000 records with
a total size of 15 \textsc{gb}), the time to upload a complete baseline
would also be proportionately smaller since, as we noted earlier,
the replication time is linear with respect to the repository's
size for both the news and email methods of replication. The news
approach actively traverses the repository, creating its own
news posts, and is therefore constrained primarily by bandwidth
to the news server or limits on posts imposed by the news server.
Email, on the other hand, passively waits
for existing email traffic and then ``hitches a ride'' to the destination
host. The SMTP approach is dependent on the site's 
daily email traffic to the host, and
a reduction in the number of records has a bigger impact
if the repository uses the email solution because fewer emails
will be needed to complete a baseline harvest of the repository.

\subsection{A Mature Repository}
Consider a mature repository with an initial size of 1 million records
averaging 100\textsc{kb} each (totaling 95 \textsc{gb} of data).  
If the repository
experiences a relatively low level of activity (10 new records (0.001\textsc{gb})
and 5 modifications (0.0005\textsc{gb})), the sender can maintain at least 3 copies
of the repository, including changes, for over 5 years using the NNTP method.
As before,
we simulate a fairly low bandwidth (10 \textsc{gb} per day max capacity). 
The column ``Mature'' in Table~\ref{simCombo}
lists the repository values and the policy factors for
both sender and receiver.
\begin{table}
	\centering
	\caption{Values used in simulations }
    \begin{tabular}{l  l  c  c  c }
		\multicolumn{5}{l}{ }\\
		\hline
		\cline{1-5}& &Active& Mature & New\\
	\cline{1-5}\multirow{4}{*}{Repository}	& $R$ & 100,000 & 1,000,000 & 1,000\\
	& $R_{\overline{s}}$ & 1 \textsc{mb}& 100 \textsc{kb}& 100 \textsc{kb}\\
	& $R_a$ & 100 & 10 & 100\\
	& $R_u$ & 400 & 5 & 20\\
	\cline{1-5}\multirow{2}{*}{Usenet} & $N_{ttl}$ & {30 \textsc{days}} & {30 \textsc{days}} & {30 \textsc{days}}\\
	& $S$ & {3 \textsc{days}} & {5 \textsc{days}}& {5 \textsc{days}} \\ 
	\hline
	\cline{1-5}\multirow{3}{*}{Email} & $c$ & 7378 & 7378 & 7378\\
	& $b$ & $-1.6$ & $-1.6$ & $-1.6$\\
	& $G$ & 1 & 1 & 1\\
	& $h({\textsc{d}})$ &  1 & 1 & 1\\
	\hline
	\hline
	\end{tabular}
	\label{simCombo}
\end{table}

\begin{figure}
  \centering
  \subfigure[The First 200 Days]{\label{fig:mature200}\includegraphics[totalheight=3.5in]{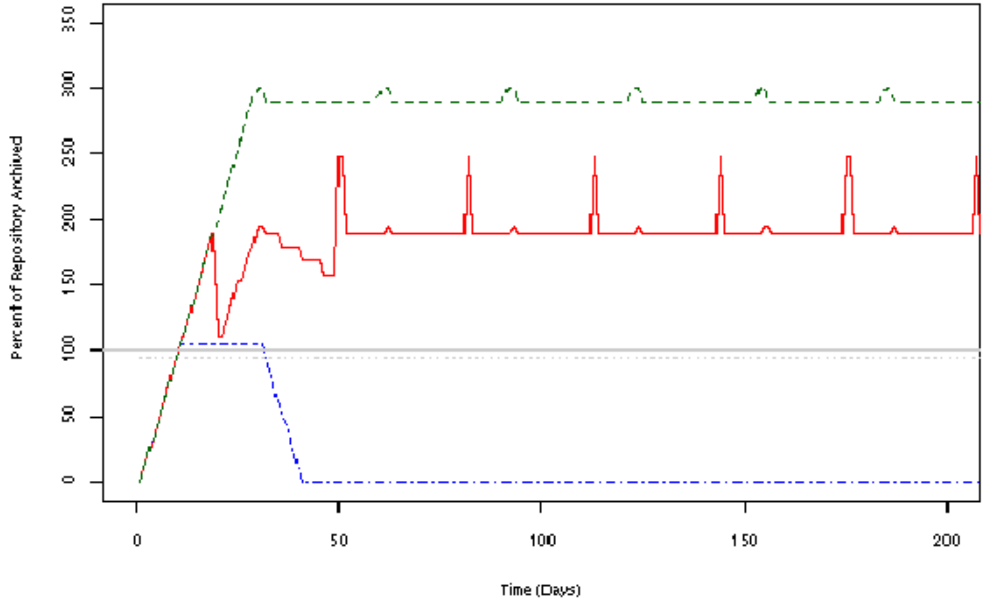}}
  \subfigure[Replication During 2000 Days]{\label{fig:mature2K}\includegraphics[totalheight=3.5in]{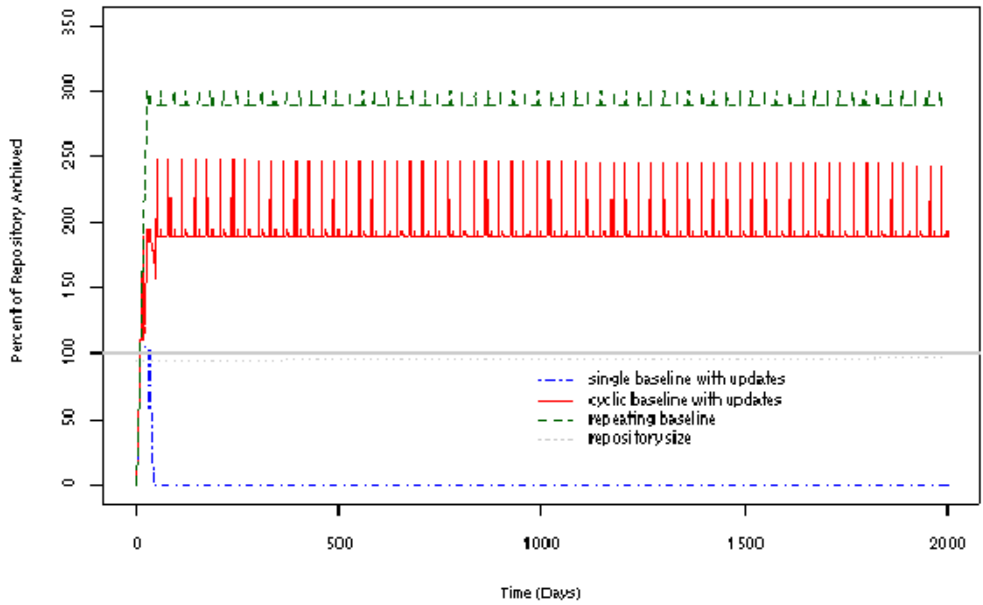}}
  \caption{Replicating a mature repository using NNTP}
  \label{fig:mature}
\end{figure}

Figure~\ref{fig:mature2K} illustrates a simulation using these values
sent to a news server with the usual 30-day expiration time. 
The single baseline policy drops off because of the deletion of records
from the news server every 30 days, 
but the cyclic and repeating baselines easily
keep up with the deletion process throughout the 2000-day simulation.
As Table~\ref{simCombo} notes, the replication target for the repeating baseline
is 3 copies, and for the cyclic baseline it is 2 copies.

\begin{figure}
  \centering
  \subfigure[Without Record History]{\label{fig:email_maturenoptr}\includegraphics[totalheight=3.5in]{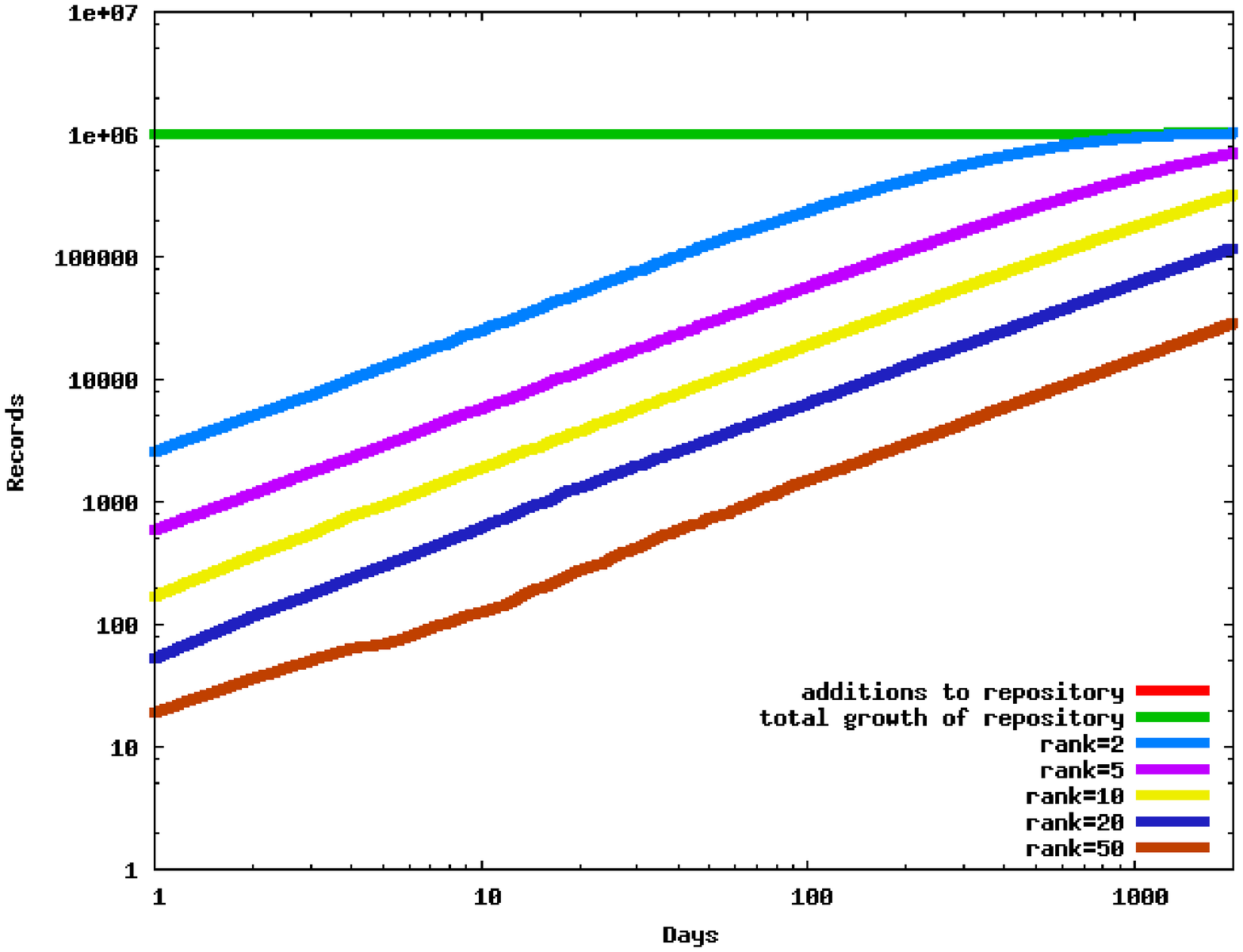}}
  \subfigure[With Record History]{\label{fig:email_matureptr}\includegraphics[totalheight=3.5in]{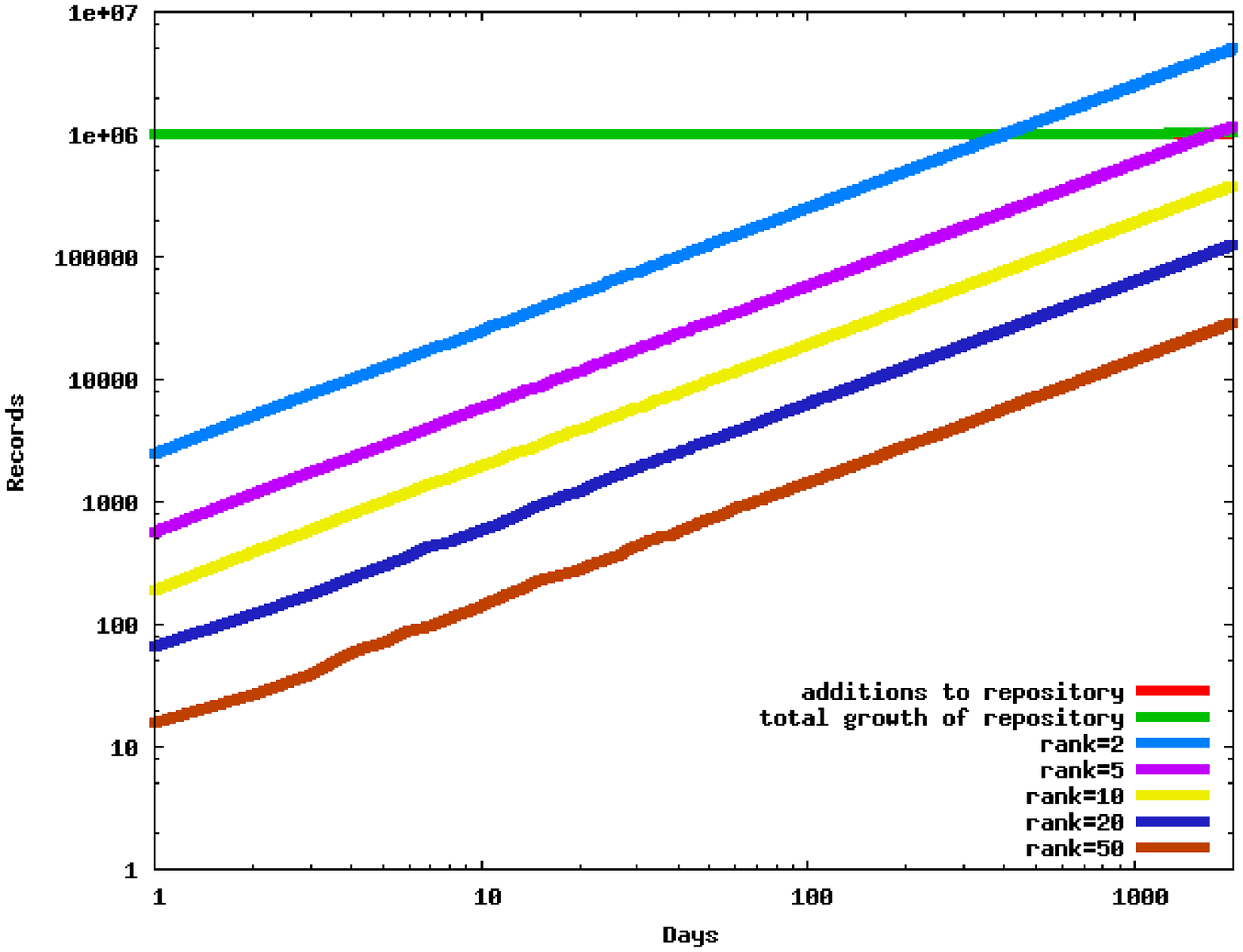}}
  \caption{Replicating a mature repository using SMTP}
  \label{fig:email_mature}
\end{figure}

Figure~\ref{fig:mature200} gives a more detailed look at the first 200
days. Notice that the cyclic baseline requires a few cycles before 
it settles down to maintaining about 2 copies on the news server. The
peaks occur because record modifications are replicated as new posts,
since previous news messages cannot be modified directly. The total
volume sent to the news server is thus the combined sum of records
and the changes to those records.
Results for the same mature repository
using the SMTP method are shown in Figure~\ref{fig:email_mature}.
We can clearly see the impact of maintaining a pointer (Figure~\ref{fig:email_matureptr})
versus without tracking the history (Figure~\ref{fig:email_maturenoptr}).

\subsection{A New, Growing Repository}
The web, of course, is full of new repositories that are fairly
active in terms of adding new content and making routine updates
every day. The column labeled ``New'' in Table~\ref{simCombo}
lists values for a hypothetical
new repository. It starts out fairly small (only 1000 records (0.1 \textsc{gb})),
but adds new records at a higher rate than the mature repository (100 records
($\sim10MB$) per day). Similarly, modifications to the new
repository happen at a similarly higher rate (20 records ($\sim2MB$) per day)
than they do in the mature repository. Although it would be reasonable
to expect the high rate of change to slow over time as the
repository matures, we maintained this high activity level 
throughout the 2000 days of the simulation.

\begin{figure}
	\centering
	\subfigure[The First 200 Days]{\label{fig:growing200}\includegraphics[totalheight=3.5in]{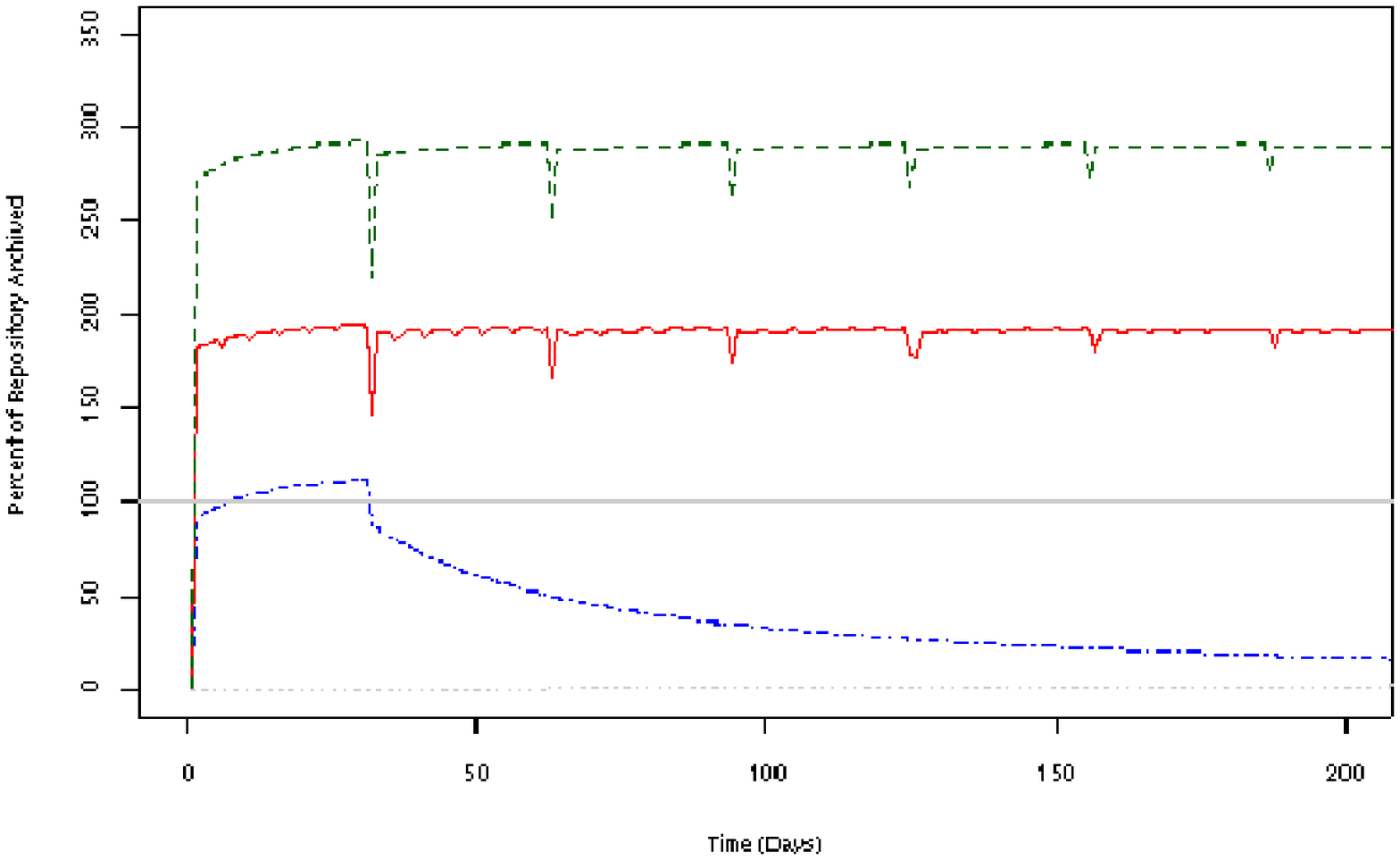}}
	\subfigure[Replication During 2000 Days]{\label{fig:growing2k}\includegraphics[totalheight=3.5in]{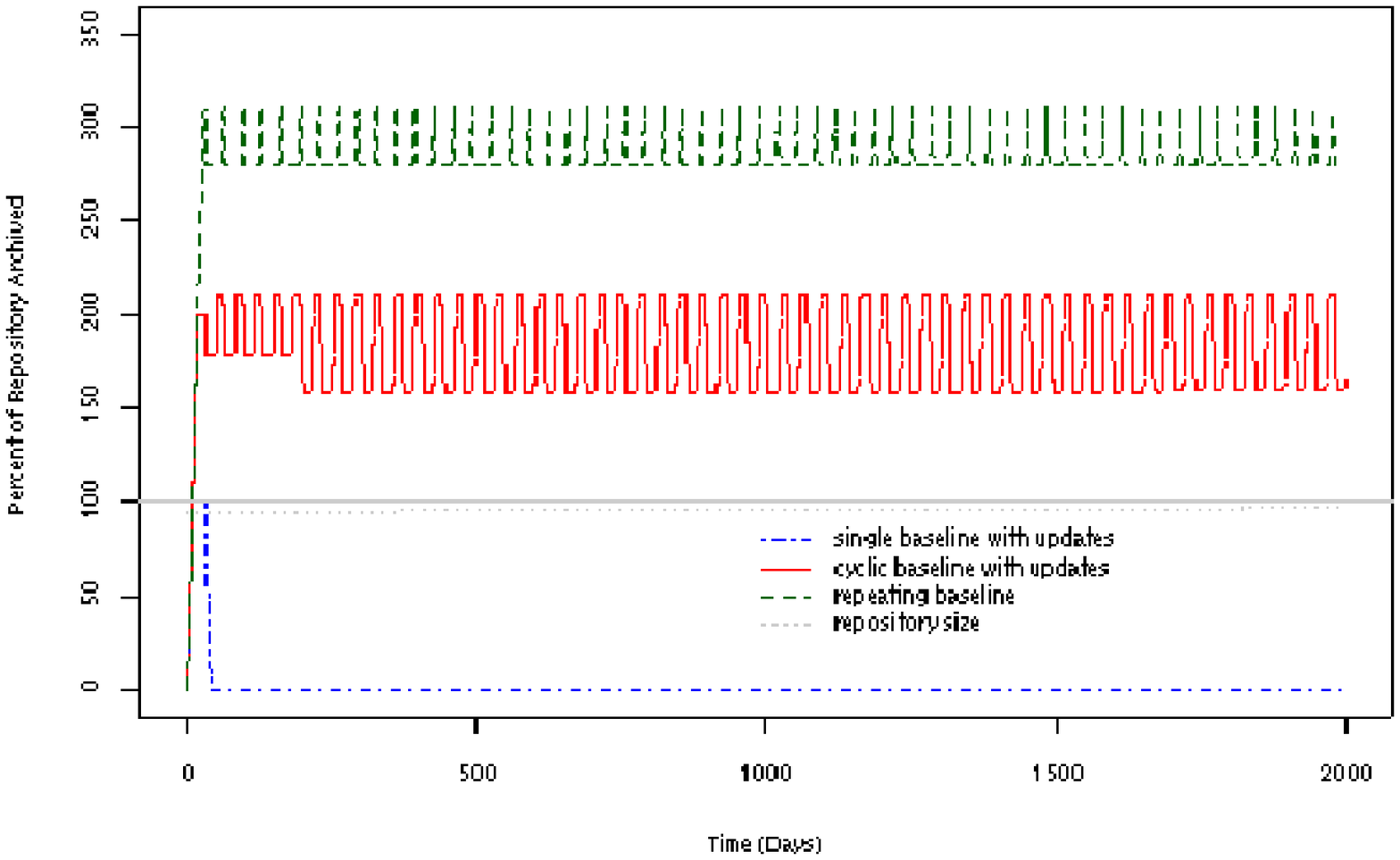}}
	\caption{Replicating a new, growing repository using NNTP}
	\label{fig:growing}
\end{figure}

Figure~\ref{fig:growing2k} shows the impact of sender policies on
maintaining replicated baselines at the news server. Despite the high
activity rate, both the cyclic baseline and the continuous baseline
policies manage to keep up with the job of replication for the entire
simulation period.  Although the news server can no longer maintain
3 full copies of the repository with the continuous baseline strategy
toward the end of the period, the news server retains at least one full
copy of the repository for the entire time frame.

\begin{figure}
  \centering
  \subfigure[Without record history]{\label{fig:email_growingnoptr}\includegraphics[totalheight=3.5in]{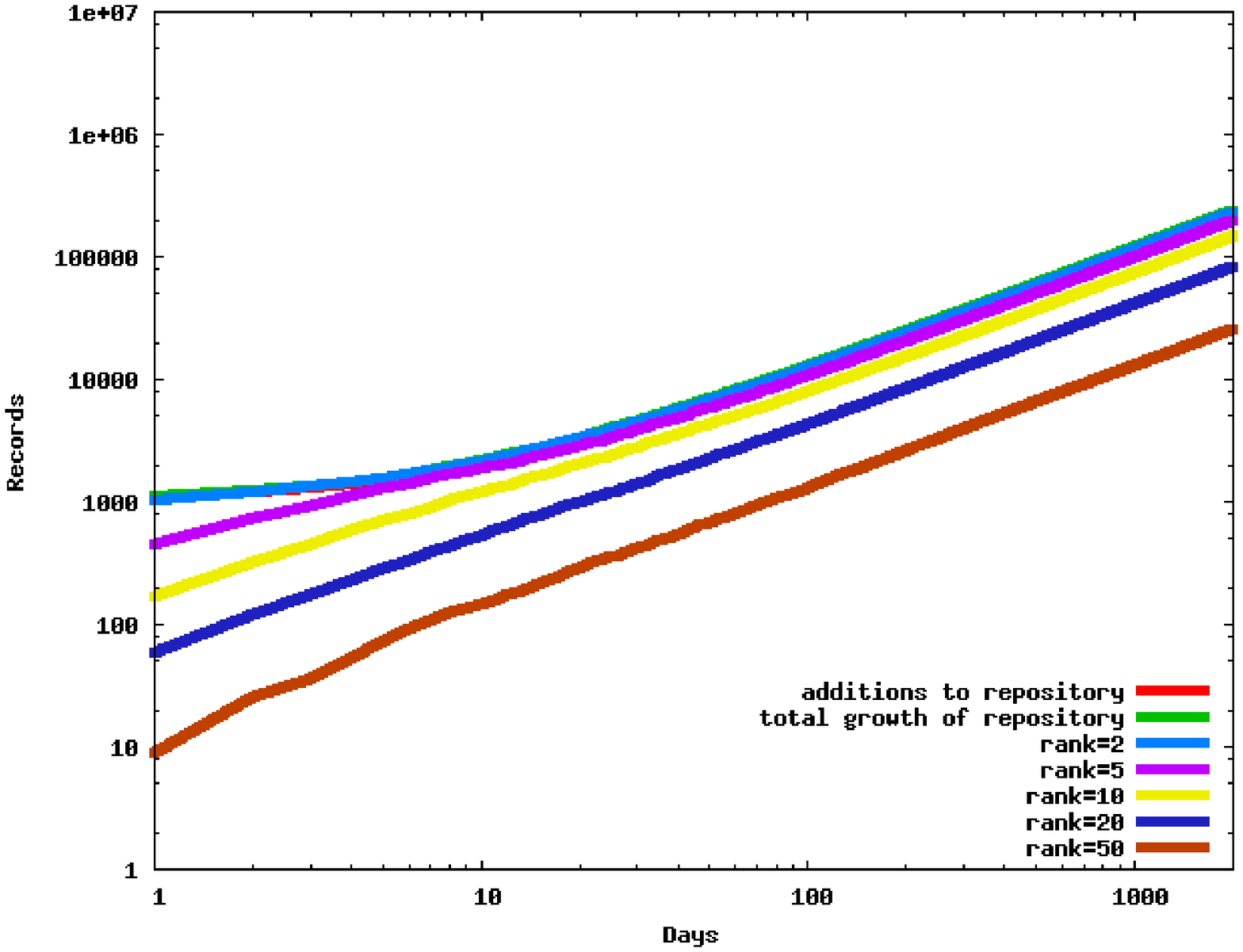}}
  \subfigure[With record history]{\label{fig:email_growingptr}\includegraphics[totalheight=3.5in]{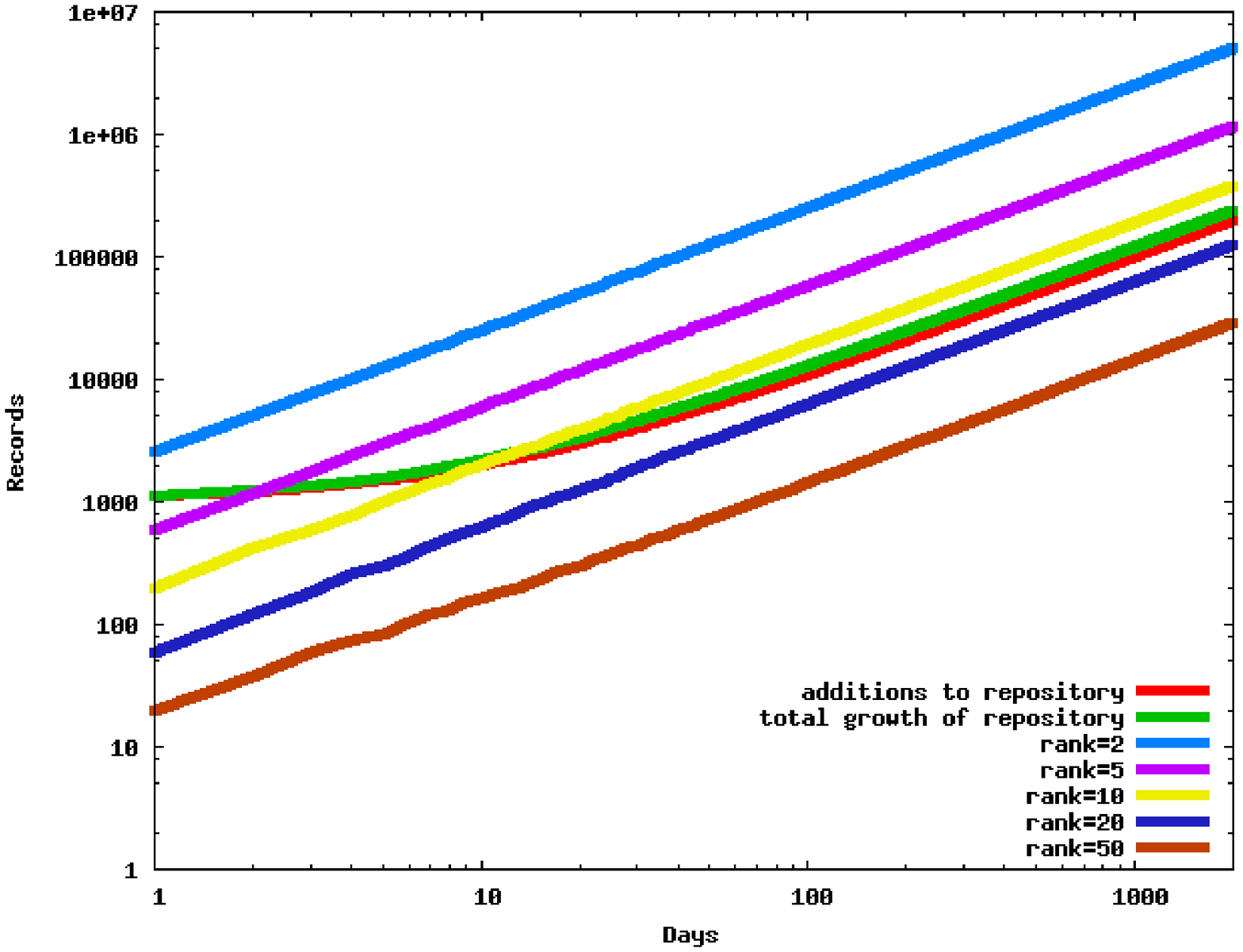}}
  \caption{Replicating a new, growing repository using SMTP}
  \label{fig:email_growing}
\end{figure}

Figure~\ref{fig:growing200} gives a closer look at the first
200 days of the simulation. The graph clearly shows the impact
of ``sleeping'' between the cyclic baselines: During the sleep
period, many new records and updates are created, and records
that were replicated earlier reach their $N_{ttl}$. This stabilizes
eventually, since even such a low bandwidth can push 10 \textsc{gb} per day
to the news server. In other words, the repository can make up
for lost time during the next ``awake'' cycle.
Compare these figures with performance for the same growing repository
using the SMTP method, as shown in Figure~\ref{fig:email_growing}. Again,
the impact of maintaining a pointer (Figure~\ref{fig:email_growingptr})
versus without tracking the history (Figure~\ref{fig:email_growingnoptr})
is obvious.

\subsection{Advertising the Repository}
One problem that repositories often face is how to improve their general
visibility to other sites and potential clients.  Buried beneath a host
of other, competing resources, repositories can become like the Dead
Sea Scrolls, hidden for digital decades.  Both the news and the email
methods of replication can help solve this problem using features unique
to OAI-PMH: Email, by virtue of disseminating information about the
repository to a wide number of hosts; news, thanks to the wide-ranging
accessibility of Usenet.  The OAI-PMH ``Identify'' response could be
effectively used to advertise the existence of a repository regardless of
the replication approach or policies. After the repository was discovered,
it could be harvested via normal means. This method can advertise even
very large repositories, since only metadata is replicated.  A simple
OAI-PMH ``Identify'' record is very small (a few kilobytes at most)
and would successfully publish the repository's existence in almost zero
time regardless of the replication approach that was used.

\section{Future Work}

Through prototypes and simulation, we have studied the feasibility
of replicating repository contents using the installed NNTP and SMTP
infrastructure.  Our initial results are promising and suggest areas
for future study.  In particular, we must explore the trade-off between
implementation simplicity and increased repository coverage.  For the SMTP
approach, this could involve the receiving email domains informing the
sender (via email) that they are receiving and processing attachments.
This would allow the sender to adjust its policies to favor those sites.
For NNTP, we would like to test varying the sending policies over time
as well as dynamically altering the time between baseline
harvests and transmission of update and additions.  
Furthermore, we plan to
revisit the structure of the objects that are transmitted, including
taking advantage of the evolving research in preparing complex digital
objects for preservation \cite{bekaert:pvdidl} \cite{vandesompel:harvest}.

\section{Conclusions}
It is unlikely that a single, superior method for digital preservation
will emerge.  Several concurrent, low-cost approaches are more likely to
increase the chances of preserving content into the future.  We believe
the piggyback methods we have explored here can be either a simple
approach to preservation, or a complement to existing methods such as
LOCKSS, especially for content unencumbered by restrictive intellectual
property rights. Even if NNTP and SMTP are not used for resource
transport, they can be effectively used for repository awareness.
We have not explored what the receiving sites do with the content once
it has been received.  In most cases, it is presumably unpacked from
its NNTP or SMTP representation and ingested into a local repository.
On the other hand, sites with apparently infinite storage capacity such
as Google Groups could function as long-term archives for the encoded
repository contents.

\section{Acknowledgments}
This work was supported by NSF Grant ISS 0455997.
We would also like to thank B. Danette Allen, who contributed to the numerical analysis.

\newpage
\appendix
\section*{Appendices}
\section{Including OAI-PMH Headers in Email and News}
\subsection{Headers in News Messages}
The actual text of the news message is formed and transmitted according to
the specification RFC 855\cite{rfc:855}. Here are the headers from an actual
message, followed by a snippet of the base64 encoded resource
(a JPEG in this case):
\begin{verbatim}
Subject:http://beatitude.cs.odu.edu:8080/modoai/10/::1155219621.6635
From:DigLib Mgr <dlmgr@beaufort.cs.odu.edu>
Date:Thu, 10 Aug 2006 14:03:45 +0000 (UTC)
Newsgroups:repository.odu.test1
Path:beatitude.cs.odu.edu!beaufort.cs.odu.edu!not-for-mail
Newsgroups:repository.odu.test1
Organization:ODU DLib
Lines:382
Message-ID:<ebfec1$rb1$1@beaufort.cs.odu.edu>
NNTP-Posting-Host:ip70-161-100-170.hr.hr.cox.net
X-Trace:beaufort.cs.odu.edu 1155218625 28001 70.161.100.170 (10 Aug 2006 14:03:45 GMT)
X-Complaints-To:news@beaufort.cs.odu.edu
NNTP-Posting-Date:Thu, 10 Aug 2006 14:03:45 +0000 (UTC)
X-Harvest_Time:2006-8-10T14:20:24Z
X-baseURL:http://beatitude.cs.odu.edu:8080/modoai/10/
X-OAI-PMH_verb:GetRecord
X-OAI-PMH_metadataPrefix:oai_didl
X-OAI-PMH_Identifier:http://beatitude.cs.odu.edu:8080/j_image.jpg
X-sourceURL:http://beatitude.cs.odu.edu:8080/modoai/10/?verb=GetRecord&i
X-sourceURL-1:dentifier=http%3A%2F%2Fbeatitude.cs.odu.edu%3A8080%2Fj_image
X-sourceURL-2:.jpg&metadataPrefix=oai_didl
X-HTTP-Header:HTTP/1.1 200 OK
Xref:beatitude.cs.odu.edu repository.odu.test1:9434

PD94bWwgdmVyc2lvbj0iMS4wIiBlbmNvZGluZz0iVVRGLTgiPz4KPE9BSS1QTUggeG1sbnM9Imh0
dHA6Ly93d3cub3BlbmFyY2hpdmVzLm9yZy9PQUkvMi4wLyIgeG1sbnM6eHNpPSJodHRwOi8vd3d3
LnczLm9yZy8yMDAxL1hNTFNjaGVtYS1pbnN0YW5jZSIgeHNpOnNjaGVtYUxvY2F0aW9uPSJodHRw
Oi8vd3d3Lm9wZW5hcmNoaXZlcy5vcmcvT0FJLzIuMC8gaHR0cDovL3d3dy5vcGVuYXJjaGl2ZXMu
b3JnL09BSS8yLjAvT0FJLVBNSC54c2QiPgo8cmVzcG9uc2VEYXRlPjIwMDYtMDgtMTBUMTQ6MTQ6
MTJaPC9yZXNwb25zZURhdGU+CjxyZXF1ZXN0IHZlcmI9IkdldFJlY29yZCIgaWRlbnRpZmllcj0i
aHR0cDovL2JlYXRpdHVkZS5jcy5vZHUuZWR1OjgwODAval9pbWFnZS5qcGciIG1ldGFkYXRhUHJl
Zml4PSJvYWlfZGlkbCI+aHR0cDovL2JlYXRpdHVkZS5jcy5vZHUuZWR1OjgwODAvbW9kb2FpLzEw
[...]
\end{verbatim}
The final section is the base64-encoded resource. We have copied only a few
lines of that portion of the message (ending with ``[\ldots]'')
since it is very long and not human-readable. 

\subsection{Headers in Email Messages}
The raw text of an email message with an appended repository record is shown below.
\begin{verbatim}
Date: Tue, 15 Aug 2006 13:09:10 -0400 (EDT)
From: martin klein <mklein@cs.odu.edu>
To: test address on then <mk@thendral.seven.research.odu.edu>
Subject: test
Message-ID: <Pine.GSO.4.58.0608151309001.7234@isis.cs.odu.edu>
MIME-Version: 1.0
Content-Type: MULTIPART/MIXED; BOUNDARY="737000039-3138878811-3085330315=:7234"
This message is in MIME format. The first part should be readable text,
   while the remaining parts are likely unreadable without MIME-aware tools.
   Send mail to mime@docserver.cac.washington.edu for more info.

X-Harvest_Time: 2006-8-15T17:9:12Z
X-baseURL: http://beatitude.cs.odu.edu:8080/modoai/
X-OAI-PMH_verb: GetRecord
X-OAI-PMH_metadataPrefix: oai_didl
X-OAI-PMH_Identifier: http://beatitude.cs.odu.edu:8080/10/pg10-8.pdf
X-sourceURL: http://beatitude.cs.odu.edu:8080/modoai/?verb=GetRecord&//
identifier=http://beatitude.cs.odu.edu:8080/10/pg10-8.pdf&metadataPrefix=oai_didl
X-HTTP-Header: HTTP/1.1 200 OK
Date: Tue, 15 Aug 2006 17:04:47 GMT
Server: Apache/2.0.49 (Fedora)
Content-Length: 6745
Connection: close
Content-Type: text/xml

--737000039-3138878811-3085330315=:7234
Content-Type: TEXT/PLAIN; charset=US-ASCII

this is a test msg

martin

perl -l -e 'print join "", reverse split //, "!nuf evah"'

--737000039-3138878811-3085330315=:7234
Content-Type: x-application/myxml; charset=US-ASCII;//
name="http://beatitude.cs.odu.edu:8080/10/pg10-8.pdf"
Content-Transfer-Encoding: BASE64
Content-Description: application/xml
Content-Disposition: attachment; filename="e247c91802a684f8fe11ccc4eab74978.xml"

PD94bWwgdmVyc2lvbj0iMS4wIiBlbmNvZGluZz0iVVRGLTgiPz4K
PE9BSS1QTUggeG1sbnM9Imh0dHA6Ly93d3cub3BlbmFyY2hpdmVzLm9yZy9PQUkvMi4wLyIgeG1s
bnM6eHNpPSJodHRwOi8vd3d3LnczLm9yZy8yMDAxL1hNTFNjaGVtYS1pbnN0YW5jZSIgeHNpOnNj
aGVtYUxvY2F0aW9uPSJodHRwOi8vd3d3Lm9wZW5hcmNoaXZlcy5vcmcvT0FJLzIuMC8gaHR0cDov
L3d3dy5vcGVuYXJjaGl2ZXMub3JnL09BSS8yLjAvT0FJLVBNSC54c2QiPgo=
PHJlc3BvbnNlRGF0ZT4yMDA2LTA4LTE1VDE3OjA0OjQ3WjwvcmVzcG9uc2VEYXRlPgo=
PHJlcXVlc3QgdmVyYj0iR2V0UmVjb3JkIiBpZGVudGlmaWVyPSJodHRwOi8vYmVhdGl0dWRlLmNz
[...]

--737000039-3138878811-3085330315=:7234--
\end{verbatim}
As with the news message sample, the base64-encoded portion of the message is 
only partially shown in this email. 

\section{Email Traffic Data} \label{app:email}
\begin{figure}
  \centering
  \subfigure[Cumulative new email receiver domains \& amount of emails]{\label{fig:cumulative_new_dom}\includegraphics[totalheight=3.4in]{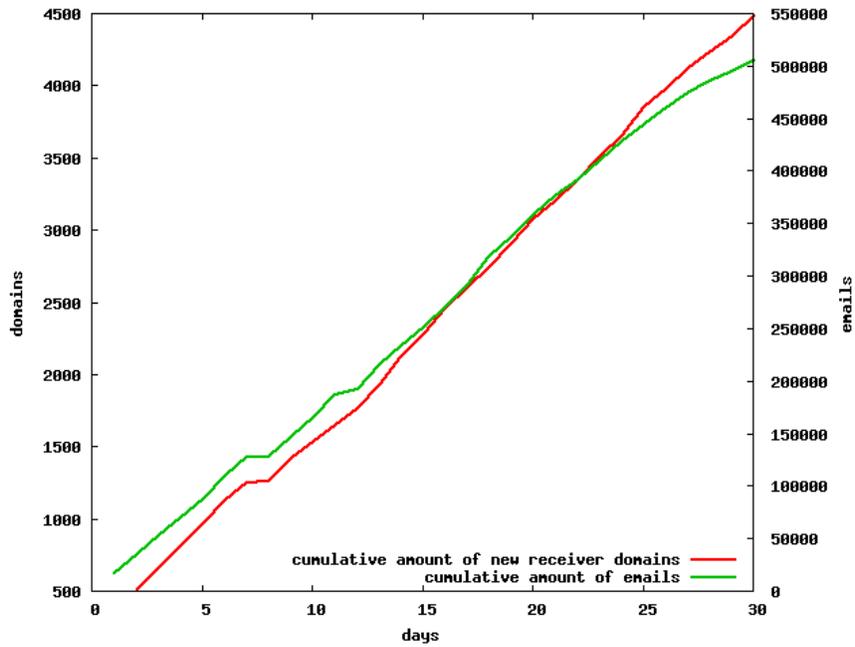}}
  \subfigure[Different receiver domains \& amount of daily email]{\label{fig:diff_rec_dom}\includegraphics[totalheight=3.4in]{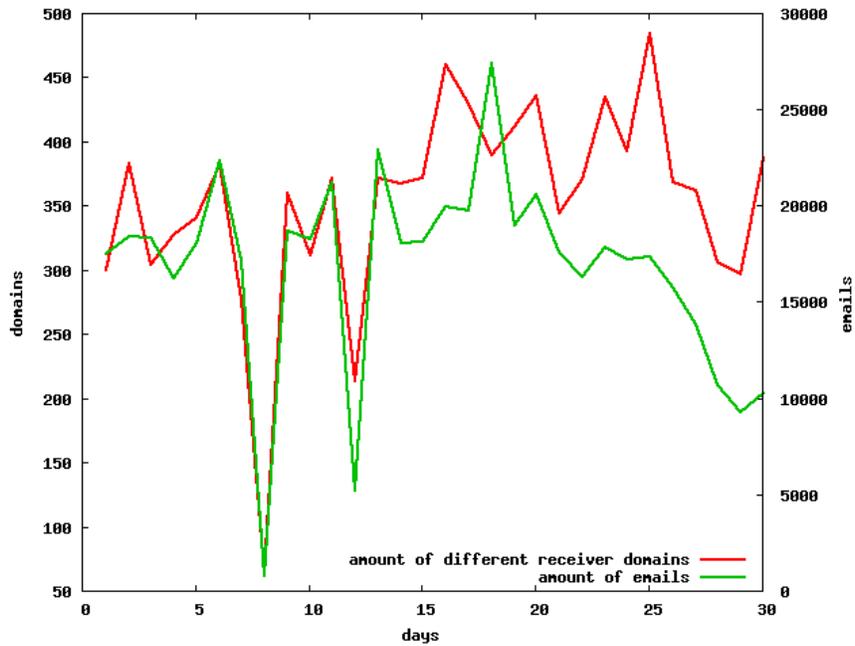}}
  \caption{Cumulative new receiver domains and amount of emails}
  \label{fig:emailRecs}
\end{figure}

In section \ref{emailProto} we analyzed the outgoing email traffic
of the Computer Science Department at Old Dominion University over a period of
30 days (January $29^{th}$ 2006 through February $27^{th}$ 2006).
Figures \ref{fig:cumulative_new_dom} and \ref{fig:diff_rec_dom}
depict the department's outbound email traffic. Note that Figure
\ref{fig:cumulative_new_dom} shows a nearly linear relationship between the 
cumulative amount of new receiver domains (scaled on the left y-axis)
and the cumulative amount of emails (the right y-axis)
sent within the observed time frame.
In figure \ref{fig:diff_rec_dom} we can see the amount of different receiver domains
per day (left y-axis) compared to the amount of emails (right y-axis) sent per day.
In both figures day one represents January $29^{th}$ and day 30 February $27^{th}$.

In Figure \ref{fig:diff_rec_dom} two dramatic
decreases in the amount of emails sent as well as in the amount of new receiver
domains are visible. Although we do not have a plausible explanation for the second low point
on Thursday, February $9^{th}$ with just 5271 outbound emails, there is a
good reason for the first, even more dramatic low point of just 828 outbound
emails on February $5^{th}$: it was Super Bowl Sunday.
These two distinctive points are also visible in Figure \ref{fig:cumulative_new_dom},
where the cumulative value for emails is close to zero compared to all
other days.

Table \ref{top_dom} shows the top 50 ranked receiver domains. The
internal email traffic is dominant followed by famous email providers
like Yahoo! and Gmail.  Ignoring internal emails (i.e., odu.edu), only 5
universities appear in the top 50, with the highest ranking university
at rank 33.  These points support the argument that email might rather
be applicable for repository advertisement than efficient repository
replication.
\begin{table}
	\centering
	\caption{Top 50 ranked receiver domains at ODU CS Department email}
	\begin{small}
    \begin{tabular}{|l|l|l|}
	\multicolumn {3}{l}{} \\
	\hline
	\bf{Rank} & \bf{Emails} & \bf{Domain} \\
	\hline
	\cline{1-3}1 & 220582 & ODU.EDU \\
        2 & 36508 & YAHOO.COM \\
        3 & 30955 & GMAIL.COM \\
        4 & 14045 & COX.NET \\
        5 & 9960 & PRADELLA.BIZ \\
        6 & 8094 & VERIZON.NET \\
        7 & 3946 & COMCAST.NET \\
        8 & 3478 & HOTMAIL.COM \\
        9 & 3238 & POBOX.COM \\
        10 & 3178 & BOUNCE.NITENIGHTPROMO.COM \\
        11 & 3164 & 0733.COM \\
        12 & 3009 & ACM.ORG \\
        13 & 2897 & BOUNCE.CHARISMADIRINC.COM \\
        14 & 2702 & BOUNCE.BLAYWAY.COM \\
        15 & 2673 & INTERNATIONALCSPEDITION.COM \\
        16 & 2617 & BOUNCE.DIRECTGAUGEBLUE.COM \\
        17 & 2555 & TAKLAM.COM \\
        18 & 2289 & LARC.NASA.GOV \\
        19 & 2042 & SPEAKEASY.NET \\
        20 & 1987 & SYSABEND.ORG. \\
        21 & 1983 & QUALCOMM.COM \\
        22 & 1968 & GLAVES.ORG \\
        23 & 1866 & BOUNCE.BLUEWATERSKY.COM \\
        24 & 1838 & CW.NET \\
        25 & 1828 & BOUNCE.TICKYTRACKY.COM \\
        26 & 1804 & CABLE.WANADOO.NL \\
        27 & 1765 & ABSOLUTEMOTION.COM \\
        28 & 1699 & NAXS.NET \\
        29 & 1643 & E-STANDARD.BIZ \\
        30 & 1642 & BOUNCE.DODGEROCKBALL.COM \\
        31 & 1633 & FUSEMAIL.COM \\
        32 & 1502 & JASONONTHE.NET \\
        33 & 1501 & CL.CAM.AC.UK \\
        34 & 1459 & COMCONNECTION.NET \\
        35 & 1441 & ABDATOS.COM \\
        36 & 1423 & AUERBACH.COM \\
        37 & 1418 & BOUNCE.SKYBEACHTIE.COM \\
        38 & 1394 & CHRISTENSENARMS.COM \\
        39 & 1358 & NCSI.IISC.ERNET.IN \\
        40 & 1347 & CWU.EDU \\
        41 & 1304 & BILLINGHAM-SL.COM \\
        42 & 1216 & BARR-MULLIN.COM \\
        43 & 1211 & EXODUS.NET \\
        44 & 1175 & IGETSMART.COM \\
        45 & 1134 & MATHS.ANU.EDU.AU \\
        46 & 1122 & BOUNCE.TUNETIMELAP.COM \\
        47 & 1098 & VIRTUA.COM.BR \\
        48 & 950 & NSWC.NAVY.MIL \\
        49 & 938 & KOLACHE.CS.TAMU.EDU \\
        50 & 936 & LIMITED-ONLINEOFFERS.COM \\
	\hline
	\end{tabular}
\end{small}
	\label{top_dom}
\end{table}
\end{document}